\documentclass[12pt,preprint]{aastex61}

\begin{document}
 
\newcommand{\kms}{km s$^{-1}\;$}
\newcommand{\msun}{M_{\odot}}
\newcommand{\rsun}{R_{\odot}}
\newcommand{\teff}{T_{\rm eff}}
\newcommand{\kep}{{\it Kepler}~}
\makeatletter
\newcommand{\Rmnum}[1]{\expandafter\@slowromancap\romannumeral #1@}
\newcommand{\rmnum}[1]{\romannumeral #1}
 
\title{The K2 M67 Study: A Curiously Young Star in an Eclipsing Binary in an Old Open Cluster
\footnote{Based on observations made at Kitt Peak National Observatory, National Optical Astronomy Observatory, which is operated by the Association of Universities for Research in Astronomy (AURA) under a cooperative agreements with the National Science Foundation; with the Tillinghast Reflector Echelle Spectrograph (TRES) on the 1.5-m Tillinghast telescope, located at the Smithsonian Astrophysical Observatory’s Fred L. Whipple Observatory on Mt. Hopkins in Arizona; the HARPS-N spectrograph on the Italian Telescopio Nazionale Galileo (TNG), operated on the island of La Palma by the INAF – Fundacion Galileo Galilei (Spanish Observatory of Roque de los Muchachos of the IAC); the Las Cumbres Observatory Global Telescope network.}}

\author[0000-0003-4070-4881]{Eric L. Sandquist}
\affiliation{San Diego State University, Department of Astronomy, San
  Diego, CA, 92182 USA}

\author[0000-0002-7130-2757]{Robert D. Mathieu}
\affiliation{University of Wisconsin-Madison, Department of
  Astronomy, Madison, WI 53706, USA}

\author[0000-0002-8964-8377]{Samuel N. Quinn}
\affiliation{Harvard-Smithsonian Center for Astrophysics,
  Cambridge, MA 02138, USA}

\author{Maxwell L. Pollack}
\affiliation{University of Wisconsin-Madison, Department of
  Astronomy, Madison, WI, 53706 USA}

\author[0000-0001-9911-7388]{David W. Latham}
\affiliation{Harvard-Smithsonian Center for Astrophysics,
  Cambridge, MA 02138, USA}

\author{Timothy M. Brown}
\affiliation{Las Cumbres Observatory Global Telescope, Goleta, CA
  93117, USA}

\author{Rebecca Esselstein}
\affiliation{University of Oxford, Keble Road, Oxford OX3 9UU, UK}

\author[0000-0003-1453-0574]{Suzanne Aigrain}
\affiliation{University of Oxford, Keble Road, Oxford OX3 9UU, UK}

\author{Hannu Parviainen}
\affiliation{University of Oxford, Keble Road, Oxford OX3 9UU, UK}

\author[0000-0001-7246-5438]{Andrew Vanderburg}
\affiliation{Harvard-Smithsonian Center for Astrophysics,
  Cambridge, MA 02138, USA}

\author[0000-0002-4879-3519]{Dennis Stello}
\affiliation{School of Physics, The University of New South Wales,
  Syndey NSW 2052, Australia}
\affiliation{Sydney Institute for Astronomy (SIfA), School of Physics,
  University of Sydney, NSW, 2006, Australia}
\affiliation{Stellar Astrophysics Centre, Department of Physics and
  Astronomy, Aarhus University, Ny Munkegade 120, DK-8000 Aarhus C,
  Denmark}

\author[0000-0002-9322-0314]{Garrett Somers}
\affiliation{Department of Physics and Astronomy, Vanderbilt
  University, 6301 Stevenson Circle, Nashville, TN 37235, USA}

\author[0000-0002-7549-7766]{Marc H. Pinsonneault}
\affiliation{Department of Astronomy, Ohio State University,
140 W. 18th Avenue, Columbus, OH 43210, USA}

\author[0000-0002-4818-7885]{Jamie Tayar}
\affiliation{Department of Astronomy, Ohio State University,
140 W. 18th Avenue, Columbus, OH 43210, USA}

\author{Jerome A. Orosz}
\affiliation{San Diego
  State University, Department of Astronomy, San Diego, CA 92182, USA}

\author{Luigi R. Bedin}
\affiliation{Istituto Nazionale Astrofisica di Padova -
  Osservatorio Astronomico di Padova, Vicolo dell'Osservatorio 5,
  I-35122 Padova, Italy}

\author{Mattia Libralato}
\affiliation{Space Telescope Science
  Institute, 3700 San Martin Drive, Baltimore, MD 21218, USA}
\affiliation{Istituto Nazionale Astrofisica di Padova -
  Osservatorio Astronomico di Padova, Vicolo dell'Osservatorio 5,
  I-35122 Padova, Italy}
\affiliation{Dipartimento di Fisica e Astronomia `Galileo
  Galilei', Universit\`{a} di Padova, Vicolo dell'Osservatorio 3,
  Padova I-35122, Italy}
  
\author{Luca Malavolta}
\affiliation{Istituto Nazionale Astrofisica di Padova -
  Osservatorio Astronomico di Padova, Vicolo dell'Osservatorio 5,
  I-35122 Padova, Italy}
\affiliation{Dipartimento di Fisica e Astronomia `Galileo
  Galilei', Universit\`{a} di Padova, Vicolo dell'Osservatorio 3,
  Padova I-35122, Italy}

\author{Domenico Nardiello}
\affiliation{Istituto Nazionale Astrofisica di Padova -
  Osservatorio Astronomico di Padova, Vicolo dell'Osservatorio 5,
  I-35122 Padova, Italy}
\affiliation{Dipartimento di Fisica e Astronomia `Galileo
  Galilei', Universit\`{a} di Padova, Vicolo dell'Osservatorio 3,
  Padova I-35122, Italy}

\correspondingauthor{Eric L. Sandquist}
\email{esandquist@mail.sdsu.edu}

\begin{abstract}
We present an analysis of a slightly eccentric ($e=0.05$), partially
eclipsing long-period ($P = 69.73$ d) main sequence binary system (WOCS 12009,
Sanders 1247) in the benchmark old
open cluster M67. Using \kep K2 and ground-based
photometry along with a large set of new and reanalyzed spectra, we
derived highly precise masses ($1.111\pm0.015$ and $0.748\pm0.005
\msun$) and radii ($1.071\pm0.008\pm0.003$ and $0.713\pm0.019\pm0.026
\rsun$, with statistical and systematic error estimates) for the
stars. The radius of the secondary star is in agreement with
theory.
The primary, however, is approximately
$15\%$ smaller than reasonable isochrones for the cluster
predict. Our best explanation is that the primary star was produced
from the merger of two stars, as this can also account for the
non-detection of photospheric lithium and its higher temperature
relative to other cluster main sequence stars at the same $V$
magnitude. To understand the dynamical characteristics (low
measured rotational line broadening of the primary star and the low
eccentricity of the current binary orbit), we believe that the most
probable (but not the only) explanation is the tidal evolution of a
close binary within a primordial triple system (possibly after a
period of Kozai-Lidov oscillations), leading to merger approximately 1
Gyr ago. This star appears to be a future blue straggler that is being
revealed as the cluster ages and the most massive main sequence stars
die out.
\end{abstract}


\section{Introduction}

Star clusters provide spectacular opportunities for testing the theory
of stellar evolution thanks to the large samples of stars that are
nearly coeval and chemically similar but heterogeneous in mass. But
while astronomers have developed and applied many techniques for
studying cluster stars in a relative sense (comparing stars in the
same cluster by minimizing uncertainties due to imperfectly measured
quantities like distance, reddening, and others), the use of precise absolute
quantities that enable direct comparisons with models is much rarer. 

The open cluster M67 is an important one for testing models --- of
all of the nearest clusters, M67 has a chemical composition and age
that is closest to that of the Sun. If our models can match the Sun's
observed properties, M67 should allow us to extend our understanding
to stars of different mass.
As part of the K2 M67 Study \citep{k2proj}, we are analyzing eclipsing
binary stars to provide a precise mass scale for the cluster stars as
an aid to these model comparisons. The K2 mission has uncovered
eclipsing binaries that would have been difficult to identify from the
ground, and has made it possible to precisely study even those systems
with shallow eclipses. Our goal here is to measure masses and radii to
precisions of better than 1\% in order to be comparable to results
from the best-measured binaries in the field
\citep{andersen,torres,debcat}.

WOCS 12009 (also known as Sanders 1247, EPIC 211409263; $\alpha_{2000}
= 08^{\mbox{h}}51^{\mbox{m}}37\fs25$, $\delta_{2000} =
+11\degr46\arcmin55\farcs7$) in M67 was first
discussed as a spectroscopic binary star by \citet{latham92}, who
identified the orbital period of 69.77 d and a small but non-zero
eccentricity ($e = 0.056 \pm 0.010$). However, it was not until
photometric observations as part of the {\it Kepler} K2 mission that
the system was identified as eclipsing. In the cluster's
color-magnitude diagram (CMD), it sits just below the turnoff, which
is an indication that the more evolved star in the binary could also
provide a means of precisely age dating the cluster. There was at
least one odd observation of the binary, however: \citet{jones} found
only an upper limit on the lithium abundance ($A$(Li)$ = \log
(N_{\mbox{Li}} / N_{\mbox{H}}) + 12 < 1.32$). Considering that most
other M67 stars of similar brightness have detectable lithium, this
was a sign that this star has somehow managed to largely burn its stores.

Because of the binary's potential significance, we sought to carefully
measure the characteristics of these stars.  In section \ref{obs}, we
describe the photometric and spectroscopic data that we collected and
analyzed for the binary. In section \ref{analy}, we describe the
modeling of the binary system. In section \ref{discuss}, we discuss
the results and the interpretation of the system.

\section{Observational Material and Data Reduction}\label{obs}

\subsection{K2 Photometry}

M67 was observed during Campaign 5 of the K2 mission. For the central
region of the cluster, all pixel data were recorded to form
``superstamp'' images for each long cadence (30 min) exposure. One
primary eclipse and two secondary eclipses were observed for WOCS
12009 during the nearly 75 d campaign, enabling the immediate
confirmation of the spectroscopic orbital period. Once the eclipsing
binary had been identified, we were able to identify observations of
earlier eclipses in other ground-based datasets, as discussed
below. On its own, the K2 light curve is particularly important to model for its
higher precision photometry.

Because of the incomplete gyroscopic stabilization during the K2
mission, systematic effects on the light curve are substantial. We
used several different light curves for WOCS 12009 as part
of our investigation of systematic error sources. Our primary light
curve choice is the one from \citet{nardiellom67}, which
employed an effective point spread function and high-resolution
imagery of the field to identify and subtract neighbors
\citep[hereafter labelled ``PSF-based approach'';][]{libralato}. Use of the \citeauthor{nardiellom67} detrended
light curve or small photometric apertures generally involved larger numbers of
discrepant data points or more complicated light variation around the
primary eclipse.  We chose to use their raw
4.5 pixel aperture 
photometry and further detrend the light
curve using a 60-point median moving boxcar filter, corresponding to a
30 hour time span. This allows a small amount of
contaminating light from other stars in the field into the aperture,
but we allow for this contamination.

We experimented with two additional K2 light curves. The first is the
K2SFF light curve derived using the method of \citet{keplerav} and
\citet{k2av} that involved stationary aperture photometry along with
correction for correlations between the telescope pointing and the
measured flux. The light curve that resulted still retained small
trends over the long term, so this was removed by fitting the
out-of-eclipse points with a low-order polynomial and subtracting the
fit. A few eclipse measurements (and others out of eclipse) were
discarded from this light curve because of thruster firings and other
events that produced discrepant points.  The second additional light curve came from
the K2SC pipeline \citep{k2ox}, which models and removes
position-dependent systematics and time-dependent trends from K2 light
curves themselves. We used a 2 pixel aperture in this light curve.

Because the K2 light curves have high signal-to-noise, they have a
substantial influence on the best-fit binary star parameter values
(see section \ref{binary}).  The scatter among the best-fit parameter
values when comparing runs with different K2 light curve reductions is sometimes
larger than the statistical uncertainties. Because there is no clearly
superior method for processing the K2 light curves, we take the
scatter in the binary model parameter values as an indicator of
systematic error resulting from the light curves.

\subsection{Ground-based Photometry}\label{gbphot}

In order to obtain color information for the binary, we obtained
eclipse photometry in $BVR_cI_c$ using several facilities. The earliest
eclipse observations we identified in archival data was made of
part of a secondary eclipse on 21 April 2003 (JD 2452750) at the Mount Laguna
Observatory (MLO) 1m telescope in $I_c$ band by one of us
(E.L.S.). After the discovery in K2 data, new observations were made at
the MLO 1m in $BVR_cI_c$ filters. The observational details are
given in Table \ref{tableobs}.  These images were processed using
standard IRAF\footnote{IRAF is distributed by the National Optical
  Astronomy Observatory, which is operated by the Association of
  Universities for Research in Astronomy, Inc., under cooperative
  agreement with the National Science Foundation.}  routines to apply
overscan corrections from each image, to subtract a master bias frame,
and to divide a master flat field frame.

Archival observations of a secondary eclipse in $V$ were found within
data from the multi-telescope campaign on M67 reported by
\citet{stellom67}. The only eclipse observations were made at the Danish 1.5m
telescope at the La Silla Observatory. To incorporate these
observations into our analysis, we matched the out-of-eclipse level in
the light curve with those of our other ground-based datasets, which
involved a $-0.004$ mag shift.

The majority of our new ground-based eclipse observations were made with
the telescopes of the Las Cumbres Observatory Global Telescope (LCOGT)
Network. We mostly used 1m telescopes, but a 0.4m was used for one
night of observation. The imagers on the 1m telescopes came in two
varieties: the SBIG ($15\farcm6 \times 15\farcm6$ field; $4096 \times
4096$ pixels; $2 \times 2$ pixel binning) and Sinistro ($26^\prime
\times 26^\prime$ field; $4096 \times 4096$ pixels)
cameras. The SBIG camera on the 0.4m had a $29^\prime \times
19^\prime$ field, and used binned $2 \times 2$ pixels. We subsequently
used images that had been processed through the standard LCOGT
pipeline in our photometric analysis below.

The brightness measurements were derived from multi-aperture photometry
using DAOPHOT \citep{daophot}. We undertook a
curve-of-growth analysis of the 12 apertures photometered per star in
order to correct all measured stars to a uniform large aperture.  We
then corrected the data for each filter to a consistent zeropoint by
using ensemble photometry \citep{honey,s1082}. This essentially uses
all measured nonvariable stars on the frame to determine magnitude
offsets resulting from differences in exposure time, airmass,
atmospheric transparency, and the like. Our implementation iteratively
fits for position-dependent corrections that result from variations in
point-spread function across the frame.  These steps each brought
noticeable reductions in the amount of scatter in the light curves.
We found that an additional step of subtracting the light curve of a
nearby bright nonvariable star also helped improve the agreement of
zeropoints from night to night. In the case of WOCS 12009, we used the
star WOCS 6008/Sanders 1242, which is about 1.3 mag brighter than WOCS
12009 and about $30\arcsec$ away on the sky.

\begin{figure}
\plotone{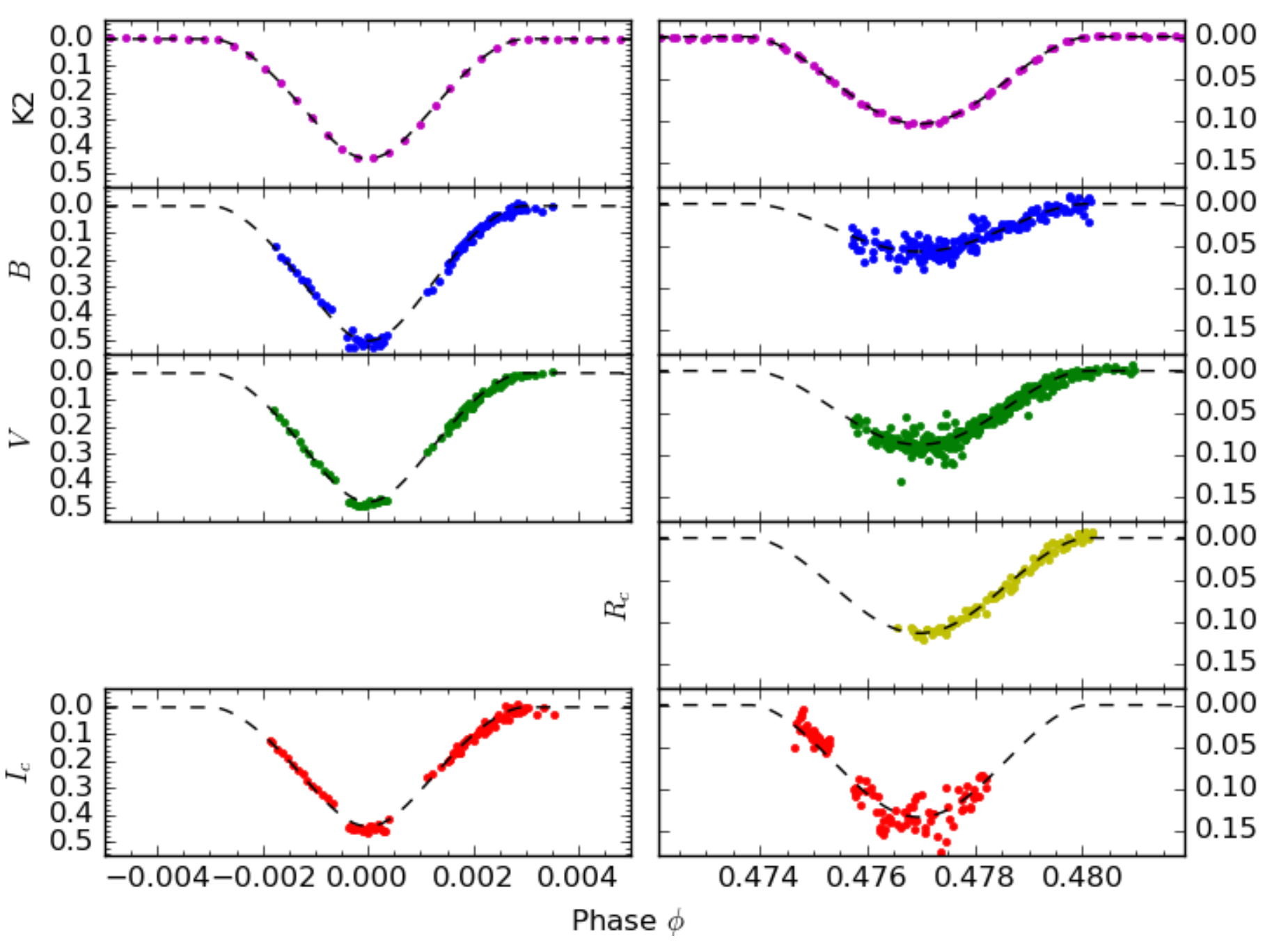}
\caption{Eclipse photometry compared with binary models. {\it Top row:} K2 photometry of the primary and secondary
  eclipse of WOCS 12009 for the PSF-based approach
  \citep{nardiellom67}. {\it Bottom rows:} Ground-based $BVR_CI_C$
  photometry of the primary and secondary eclipses. \label{phot}}
\end{figure}

After the initial reductions, we found that there were significant
color-dependent residuals in the light curves from different stars
that resulted from the use of different telescope/camera/filter
combinations. These were identified and characterized as linear
relations using non-variable stars in the field, but were only judged
to be significant in $B$ and $I_c$ filters.
The correction relations were used to correct all light curves to the
system of the MLO filter set as part of the final ensemble photometry
analysis. The specific corrections that were applied to WOCS 12009
light curves were less than 0.008 mag for all observational set-ups
except one in $I_c$ (0.014 mag).
The eclipse observations are shown in Fig. \ref{phot}, and the light curves are provided in tables in the electronic version.

\subsection{Spectroscopy}\label{spectro}

Radial velocities of M67 stars have been monitored over a very long
period, and there is a large database of WOCS 12009 observations from
different telescope / spectrograph combinations spanning almost 29
years. The earliest 23 observations we use here employed the CfA
Digital Speedometers (DS; \citealt{latham85,latham92}), which covered
516.7 to 521 nm around the Mg I b triplet.  Additional observations
were taken on the 1.5 m Tillinghast reflector at Whipple Observatory
using the Tillinghast Reflector Echelle Spectrograph (TRES; \citealt{tresspec}). TRES
spectra contained 51 echelle orders covering the whole optical range
and into the near-infrared (from about 386 to 910 nm).

Ten spectroscopic observations were taken as part of the WIYN Open
Cluster Survey (WOCS; Mathieu 2000) using the WIYN\footnote{The WIYN
  Observatory is a joint facility of the University of
  Wisconsin-Madison, Indiana University, the National Optical
  Astronomy Observatory, and the University of Missouri.} 3.5 m
telescope on Kitt Peak and Hydra multi-object spectrograph instrument
(MOS), which is a fiber-fed spectrograph capable of obtaining about 80
spectra simultaneously ($\sim$10 fibers set aside for sky measurements
and $\sim$70 for stellar spectra). Observations used the echelle
grating with a spectral resolution of 20,000 (velocity resolution
$\sim$ 15 km s$^{-1}$).  The spectra are centered at 513 nm with a 25
nm range that covers an array of narrow absorption lines around the Mg
I b triplet. Spectroscopic observations for WOCS 12009 were completed
using total integrations of either one or two hours per visit that
were split into three sub-integrations to allow for rejection of
cosmic rays.

Image processing was done using standard spectroscopic IRAF
procedures. First, science images were bias and sky subtracted, then the
extracted spectra were flat fielded, throughput corrected, and
dispersion corrected.  The spectra were calibrated using one 100 s
flat field and two bracketing 300 s ThAr emission lamp spectra for
separate integration times. A more detailed explanation of the
acquisition and reduction of the CfA DS, TRES, and WIYN Hydra
observations can be found in \citet{geller}.

We also collected two new spectra from the HARPS-N spectrograph
\citep{harpsn} on the 3.6 m Telescopio Nazionale Galileo
(TNG). HARPS-N is a fiber-fed echelle that has a spectral resolution
$R = 115000$ covering wavelengths from 383 to 693 nm. The spectra were
processed, extracted, and calibrated using the Data Reduction Software
(version 3.7) provided with the instrument.

We obtained archival spectra from two additional sources.  Eight
archived $H$-band infrared ($1.51 - 1.70 \mu$m) spectra with a
spectral resolution $R \sim 22500$ for WOCS 12009 were available from
the APOGEE project \citep{apogee} on the 2.5 m Sloan Foundation
Telescope.  Before making our radial velocity measurements, we used
APOGEE flags to mask out portions of the spectrum that were strongly
affected by sky features, and we continuum normalized the spectra
using a median filter.  \citet{pasquini} tabulated three radial
velocity measurements for the primary star as part of their survey of
solar analogs in M67, based on observations from February 2007 using
the FLAMES/GIRAFFE spectrograph at the VLT. Their spectra covered 644
to 682 nm, including the H$\alpha$ line, and we obtained the spectra
through the ESO Phase 3 Data Archive.  We continuum normalized the
spectra that had been processed and calibrated through the ESO
pipeline.

The radial velocities were derived using a spectral disentangling code
following the algorithm described in \citet{gonz}.  In each iteration
step the spectrum for each component is isolated by aligning the
observed spectra using the measured radial velocities for that
component and then averaging. This immediately de-emphasizes the lines
of the non-aligned stellar component, and after the first
determination of the average spectrum for each star, the contribution
of the non-aligned star is subtracted before the averaging to improve
the spectral separation. The radial velocities can also be remeasured
using the spectra with one component subtracted. This procedure is
repeated for both components and continued until a convergence
criterion is met.  We use the broadening function (BF) formalism
\citep{rucinski02} to measure the radial velocities using synthetic
spectra from the grid of
\citet{coelho05} as templates. BFs can significantly improve the
accuracy of radial velocity measurements in cases when the lines from
the two stars are moderately blended \citep{rucinski02}.  In the case
of the WOCS 12009 binary, the spectra of the two stars have
significant differences, so that lines of each star can usually be identified
even when the velocity separation is less than 10 km s$^{-1}$.  As a
result, we can derive reasonably reliable velocities for the two stars
at small velocity separations when we couple the disentangling
technique (and especially the subtraction of the averaged spectrum of
each star) with BFs.

The disentangling procedure requires spectra covering the same
wavelength range taken at very different orbital phases, and so we had
to work with different subsets of our spectra.  The CfA DS spectra
covered a wide range of orbital phases, but had a limited wavelength
range, and so we disentangled these spectra separately. Many of the
spectra had comparatively low signal-to-noise that made it impossible
to identify the secondary component. We used all spectra in
determining the average spectrum for the primary, but only used the 8
spectra with the most clearly detected secondary component to
determine its average spectrum. We were only able to measure reliable
velocities for the secondary star from these spectra and two others.

We disentangled the TRES and HARPS-N spectra together because they
covered similar wavelength ranges at similar resolution, and because
it would have been difficult to disentangle the two HARPS-N spectra on
their own. Because the TRES and HARPS-N spectra covered the entire
optical range, we decided to disentangle them in the wavelength ranges
of individual TRES echelle orders. We selected 14 orders with a large
number of strong lines, avoiding the bluest and reddest ends of the
optical range (due to low signal on the blue end and telluric lines on
the red end). This had the additional benefits that we could determine
the stellar velocities as means of the velocities derived from the
different orders and also get uncertainty estimates from the standard
error of the mean. These uncertainties were typically around 0.08 \kms
for the primary star and 0.4 \kms for the secondary.

The WIYN spectra were disentangled on their own because we were able
to obtain spectra covering a sufficient set of different phases,
although we had to take some care in the selection of spectra used to
calculate the averaged spectra for the two stars. We used 8 of the 10
spectra for the brighter primary star, and 6 of the best for the
secondary. Uncertainties were initially derived from the BF fits, and
were scaled to produce a reduced $\chi^2$ values of 1 based on a best
Keplerian fit to the WIYN radial velocities alone. Typical
uncertainties were around 0.15 \kms for the primary star and 1 \kms
for the secondary star.

The APOGEE spectra also had to be disentangled on their own because their
spectral range did not overlap with the other datasets. The
uncertainty estimate for each APOGEE velocity was generated from the
rms of the velocities derived from the three wavelength bands in the
APOGEE spectra. This was typically around 0.2 \kms for the primary
star and 0.7 \kms for the secondary star. The velocity was corrected
to the barycentric system using values calculated by the APOGEE
pipeline. We compared our measured velocities for the primary star with
those of the APOGEE pipeline, and we found a phase-dependent
difference with a sign consistent with the idea that the tabulated
APOGEE velocities were pulled slightly ($0.5-1.0$ km s$^{-1}$) by the
presence of the unmodeled secondary star.

We also ended up disentangling the three FLAMES/GIRAFFE spectra
\citep{pasquini} separately. For two of the three spectra we clearly
detected and resolved BF peaks for the secondary star on opposite
sides of the primary star peak. For the third spectrum, the BF
peaks were moderately blended. However, because the two
unblended spectra were at very different orbital phases, we were
successfully able to disentangle the spectra in this set and were able
to derive velocities for both stars in all three spectra. Reliable
uncertainties could not be derived from scatter around a best-fit
solution for these measurements on their own, however. As a result, we
derived uncertainty estimates from their scatter around the best-fit
to the entire velocity dataset.

Finally, we employed one radial velocity measurement from
\citet{jones} made in March 1996 using the HIRES spectrograph at the
Keck 10m telescope. Because the secondary star was not subtracted from
the spectrum before their measurement, it is possible there is a
small bias in the direction of the system velocity for that
measurement.

Corrections were applied to the velocities from different
spectroscopic datasets. For the CfA DS velocities, we used run-by-run
corrections derived from observations of velocity standards.
Run-by-run offsets measured for the TRES spectra were no more than 17
m s$^{-1}$, and although we did apply them, they are much smaller than
the measurement uncertainties.
Velocity offsets for individual Hydra fibers in the WIYN dataset were
determined and applied to the WOCS 12009 spectra, although these were
of different sign and no more than about $\pm0.2$ \kms in
general.

Because the synthetic spectra used in disentangling do not account for
gravitational redshifts of the stars, our velocities should have an
offset relative to velocities derived using an observed solar
template. For consistency with the large survey of \citet{geller}, we
subtracted the combined gravitational redshift of the Sun and
gravitational blueshift due to the Earth (0.62 km s$^{-1}$) from the
velocities we tabulated. We did not apply the gravitational redshift
correction to the \citeauthor{jones} Keck HIRES velocity because it
was derived using a solar template. Based on the derived masses and
radii for the stars in the binary, they should have slightly larger
gravitational redshifts than the Sun (0.66 and 0.69 km s$^{-1}$,
versus 0.64 \kms for the Sun) but we did not correct for these small
differences.  We did, however, allow for different system velocities for
the two stars in our fits to allow for effects like this and convective
blueshifting that could otherwise affect the measured radial velocity amplitudes
$K_1$ and $K_2$ (and the stellar masses) if not modeled.  The
measured radial velocities are presented in Table \ref{spectab}.

\begin{figure}
\plotone{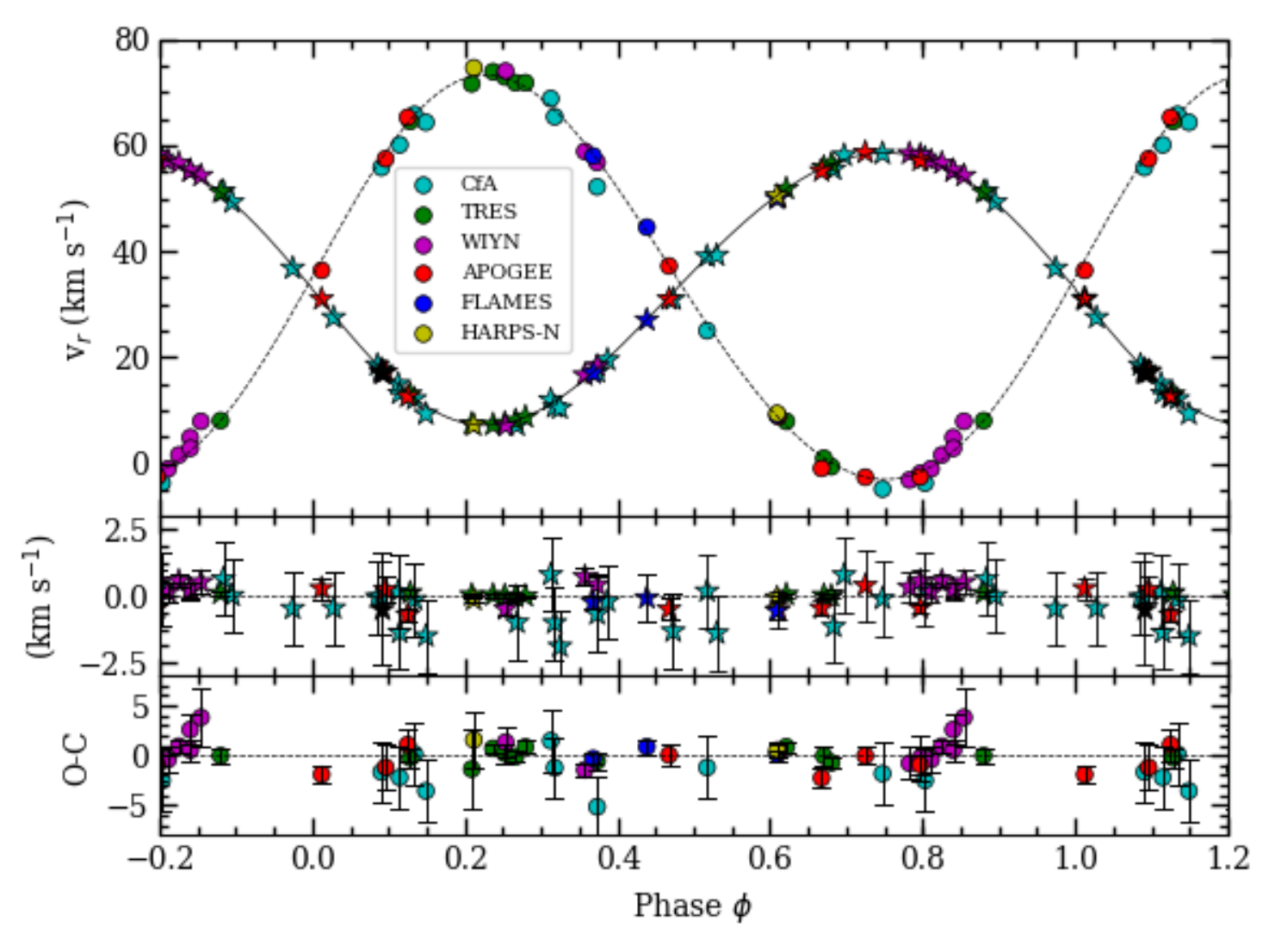}
\caption{Phased radial velocities for WOCS 12009, along with the best
  fit model (using the analytic limb darkening law). Measurements for
  the primary star are shown with star symbols and with
  circles for the secondary star. The lower panels show the observed
  minus computed values with error bars scaled to give a reduced
  $\chi^2=1$ (see \S \ref{binary}).\label{rvplot}}
\end{figure}

Even with these corrections to the velocities, we are attentive to the
possibility that systematic differences might exist from dataset to
dataset.  The phased radial velocity measurements are plotted in
Fig. \ref{rvplot}, showing the best fit to the orbit using all
measurements. Among the measured primary star velocities, mean
residuals (observed minus computed) were fairly small: 0.18 \kms for
WIYN, 0.05 \kms for TRES, $-0.07$ \kms for HARPS-N, $-0.23$ \kms for
APOGEE and FLAMES, and $-0.45$ \kms for CfA DS. Because we do not see
clear evidence of systematic shifts between spectroscopic datasets, we
have not applied any corrections of this sort. We have opted not to
scale the uncertainties on the datasets individually, but instead
scale uncertainties on all velocity measurements for each star to
separately produce a reduced $\chi^2$ measurement of 1. This preserves
information on the relative precision of measurement in each dataset
(which affects their weighting for the final fits) that was most often
determined {\it a posteriori}. The primary star uncertainties were
scaled upward by a factor of 2.1 and the secondary star uncertainties
were scaled by a factor of 1.38.

For the purpose of later discussions of the reliability of the stellar
mass and radius measurements, we summarize here that we have consistent radial velocity
information on both stars from five different spectrographs in a
variety of optical and infrared wavelength ranges. This gives us confidence
that systematic errors in the radial velocities do not
significantly affect the masses and radii we will derive.

\section{Analysis}\label{analy}

\subsection{Cluster Membership}

There have been many studies of M67 that have included proper motion
measurements and estimations of membership probabilities. All of the
studies agree that WOCS 12009 is a highly-probable member. The lowest
published membership probability (71\%) comes from \citet{zhaopm}, but
all others are greater than 93\%
\citep{sanderspm,girardpm,yadav,kkpm,nardiello}. In our later use of
proper motion membership probabilities to produce a color-magnitude
diagram cleaned of field stars, we made use of all of these sources,
computing a straight average of all of the probabilities published for
each star to assign a final probability.

\citet{geller} also published membership probabilities based on their
radial velocity survey, and again WOCS 12009 was assigned high
probability of membership (98\%). Our modeling of the radial
velocities presented below indicates that both stars in the binary
have high membership probabilities based on the system velocities we
derive: $\gamma_1 = 33.81$ \kms and $\gamma_2 = 33.75$ \kms.
\citeauthor{geller} found the mean cluster velocity to be
33.64 \kms with a radial velocity dispersion of $0.59^{+0.07}_{-0.06}$
km s$^{-1}$.

Based on all of the kinematic evidence, the binary is a high
probability cluster member. Additional evidence comes from the
distance moduli derived from the radius, $\teff$, and $V$ magnitude of
the stars (see \S \ref{dm}), and from the position of the components of the
binary in the CMD once the light from the two stars has been
disentangled (see \S \ref{cmd}).

\subsection{External Constraints on the Binary}\label{constrain}

Surface temperature is somewhat important for constraining the limb
darkening coefficients to be used in the eclipse modeling (section
\ref{binary}), but is even more important for the distance modulus
determination in \S \ref{dm}.

Photometric determinations of $\teff$ using the total system
photometry will be affected by the contributions of the secondary
star, but the calculation will be a lower limit to the temperature of
the primary star and will be closer to the actual temperature the larger the fraction of
the system light the primary contributes. \citet{jones} found 5889 K
from $BV$ photometry. Using the Stromgren photometry of
\citet{balaguer} and the $\teff$ calibration of \citet{onehag}, we
calculate 5920 K from the $(b-y)$ color and a reddening
$E(b-y)=0.030$. The APOGEE pipeline provides a temperature of $5950
\pm 70$ K from its best fit to its infrared $H$-band spectra using
synthetic templates, but again this will be somewhat affected by the
light of the secondary star, which is more prominent in the infrared.

With luminosity ratios determined from the binary star analysis below,
we are able to separate the light of the two stars and estimate
temperatures for them. For this purpose, we used photometry from
\citet{mephot} in the optical
along with the color-$\teff$ relations of \citet{casagrande}. Assuming
a reddening $E(B-V)=0.041\pm0.004$ \citep{taylor}, the weighted
averages of values computed from ($B-V$) and ($V-I_C$)
were 6110
K and 4760
K for the primary and secondary stars,
respectively.  These values are in good agreement with temperatures in
the synthetic spectra that optimized our detection of broadening
function signal for radial velocity measurement (6000 and 6250 K for
the primary, and 4500 and 4750 K for the secondary).

\subsection{Binary Star Modeling}\label{binary}

To simultaneously model the ground-based radial velocities and
photometry and \kep K2 photometry, we used the ELC code
\citep{elc}. We assumed the stars are spherical, and we did not
attempt to model spots.
The core of the model fit involved a set of 14 parameters: orbital
period $P$, time of conjunction (primary eclipse) $t_c$, velocity
semi-amplitude of the primary star $K_1$, mass ratio $q=M_2 /
M_1=K_1/K_2$, systematic radial velocities\footnote{We allow for the
  possibility of differences for the two stars that could result from
  differences in convective blueshifts or gravitational redshifts.}
$\gamma_1$ and $\gamma_2$, the combinations of eccentricity ($e$) and
argument of periastron ($\omega$) $e \cos \omega$ and $e \sin \omega$,
inclination $i$, ratio of the primary radius to average orbital
separation $R_1 / a$, ratio of radii $R_1 / R_2$, temperature of the
primary $T_1$, temperature ratio $T_2 / T_1$, and contamination of the
K2 light curve by other sources (independent of the $BVR_CI_C$ light
curves).  The K2 contamination parameter is used to account for
possible dilution of the light curves by stars that are not physically
associated with the binary. \kep pixels are large by the standards of
most ground-based cameras, and in the M67 cluster field, light
contamination is a greater concern than in the majority of the field
of view. The two stars of greatest significance here are II-216
($V=19.45$)
and WOCS 6008/Sanders 1242 ($V=12.33$), but
their distances ($23\arcsec$ and $28\farcs6$, respectively) will mean
that their contributions are relatively small.  Because the K2 light
curves use long cadence data with 30 minute exposures, they also
effectively measure the average flux during a 30 minute exposure.  To
account for this, the ELC code integrates the computed light curves in
each observed exposure window.

Limb darkening is a continuing uncertainty in binary star fits, and to
model this we fit for 18 additional parameters using the
\citet{kipping} algorithm: two coefficients in the quadratic limb
darkening law for each star in the \kep and $BVI_C$ filter bands (and
two coefficients for the secondary star in $R_C$ because to date we
only have secondary eclipse observations). Briefly, Kipping's
algorithm has the advantages that it reduces correlation between limb
darkening law parameters, and allows for a straightforward sampling of
physically realistic values. We modified the
\citet{kipping} algorithm by restricting the second parameter $u_2$ to values
  less than 0.5 in order to force the limb darkening law to be concave
  down as it approaches the limb. By systematically exploring the
parameter space, we can include our limb darkening uncertainties in
our uncertainties of the stellar parameters.

The quality of the model fit was quantified by an overall $\chi^2$. We
applied the temperature of the primary star $T_1 = 6100 \pm 100$ K
(see \S \ref{constrain}) as an observed constraint, allowing the value
to vary but applying a $\chi^2$ penalty based on deviation from the
determined value.
Because the photometric and spectroscopic datasets were pulled
together from a variety of sources, the relative weights of different
data influences the determination of $\chi^2$. For radial velocity
measurements, we utilized empirical estimates of measurement
uncertainties as much as possible, as described in section
\ref{spectro} for each spectroscopic dataset. These empirical
measurements were then scaled separately to give a reduced $\chi^2$ of
1 for the primary and secondary star datasets. Although some binary
star parameters are derived almost exclusively from the velocities,
others (like eccentricity $e$ and argument of periastron $\omega$) are
affected by the photometry as well, so it is important to try to
weight the velocity measurements properly with respect to the
photometry. For photometry, we scaled all of the {\it a priori}
measurement uncertainties upward by the same factor in order to
produce a reduced $\chi^2$ value of 1 based on separate best fits
  to data for each filter. Once this was done, we computed
simultaneous fits to the photometry and velocity datasets.

The model parameter space was sampled using an implementation of
  the parallel {\tt emcee} algorithm \citep{emcee} for Markov chain
  Monte Carlo. This algorithm allows for efficient sampling of
  alternate parameter sets using a group of ``walkers'' (in our case,
  1000) that move in parameter space around the values providing the
  best fit to the data. A proposal for a jump in the position of one
  walker is taken along a line connecting to, but extending beyond,
  another randomly chosen walker. The proposed position is accepted
  with a probability that decreases as $\chi^2$ increases. We found
  that the walkers converged to stable distributions within 2000
  steps, and these steps were discarded before forming posterior
  probability distributions from chains of 6000 step length, skipping
  every 15 steps. Approximate $1\sigma$ parameter uncertainties were
  derived from the parts of the posterior distributions containing
  68.2\% of the remaining models from the Markov chains.
The results of the parameter fits are provided in Table \ref{parmtab}.
In the last column we provide the consensus parameter values for the
most important quantities, including a statistical error estimate
derived from the results of models in and around the best fit, as well
as a systematic error estimate derived from the scatter among the runs
with the three different K2 light curves. When there was disagreement,
the K2SFF results were generally the outliers. This was particularly
noticeable in the luminosity ratios and in the secondary star radius
and gravity. We attribute this mostly to the slightly deeper eclipses
in the K2SFF light curve, and to a lesser extent to clearly discrepant
points during the one observed primary eclipse that had to be deleted.
For the radii and gravities, the combined values were derived from an
average of the results for all three K2 light curves, while we
averaged the results for the PSF-based approach and K2SC light curves for the
luminosity ratios. The systematic error quotes reflect these differences.

\section{Discussion}\label{discuss}

For good comparisons of our stellar observations with models, we need
to make some assumptions about the chemical composition. The
abundances of M67 stars have been studied extensively, and a full
review is well beyond what we need here. An important
ingredient is abundances measured relative to the Sun,
and so here we highlight studies that focused on high-precision
differential measurements using Sun-like stars as much as possible.
\citet{pace08} studied 6 solar analogs in M67 (and a solar spectrum) to derive
[Fe/H]$= +0.03\pm0.04$ (standard deviation), and 
\citet{santos} gets $+0.01\pm0.04$ from the same 6 dwarf spectra.
\citet{onehag11} did a strictly differential comparison of the solar
twin Sanders 770, and obtained [Fe/H]$= +0.023\pm0.015$. \citet{liu16}
conducted a strictly differential comparison of two solar twins in M67
(including a high resolution solar spectrum reflected from the
asteroid Hebe that was taken with the same set-up),
finding [Fe/H]=$-0.005\pm0.010$ for Sanders 770 and $-0.061\pm0.014$
for Sanders 1462. Those authors interpreted the difference in [Fe/H]
between the two stars as significant.

\citet{onehag14} did a differential comparison of 14 M67 stars on the
main sequence, turnoff, and subgiant branch relative to the solar twin
Sanders 770. Those authors find that M67 stars have abundance patterns
very similar to that of the Sun, and perhaps even more so than solar
twin candidates in the solar neighborhood. Their averaged abundances
for all their measured heavy elements except Li was
[X/H]$=-0.017\pm0.003$. They also found that the main sequence and
turnoff stars in their sample were slightly depleted in heavy elements
compared to subgiants, possibly due to the action of diffusion.

In a differential sense, M67 photospheric metallicities agree with the
Sun's within about $\pm0.03$ dex, although diffusion may affect the
abundances for a subset of M67 stars at similar level. It should be
remembered, however, that the bulk metal content of the Sun is still
uncertain \citep[e.g.][]{asplund,caffau}, and will remain an obstacle to a comprehensive analysis of M67 stars.

\subsection{Mass and Radius}\label{mrsec}

The masses and radii of the eclipsing stars are important because they are
high-precision, distance- and reddening-independent, measurable
characteristics. In the case of M67, these kinds of measurements can
potentially provide strong constraints on the cluster age.

\begin{figure}
\includegraphics[scale=0.8]{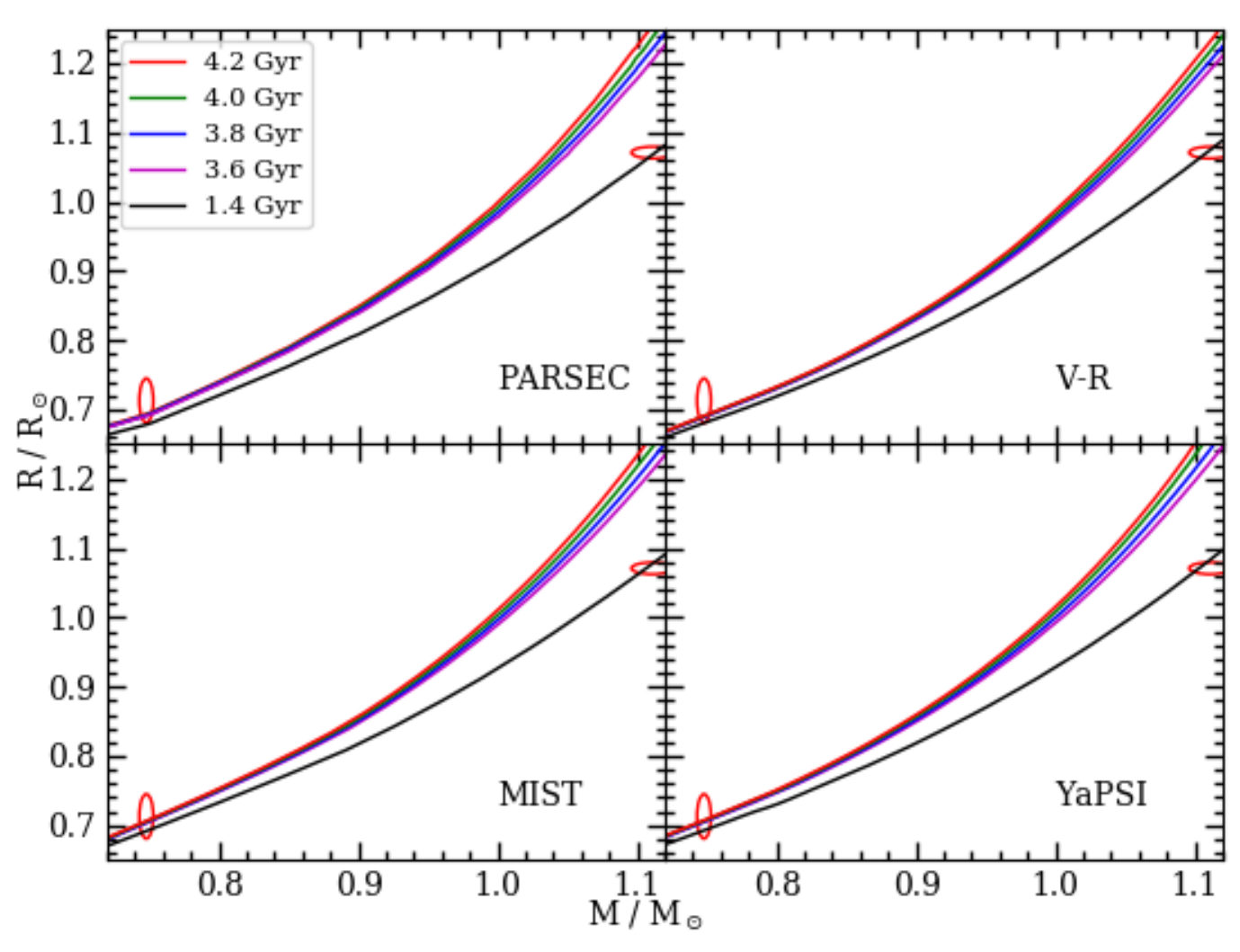}
  \caption{Mass-radius plot for the members of S1247 with $2\sigma$
  uncertainties indicated by the error ellipses. Models use $Z =
  0.0152$, 0.0188, 0.0142, and 0.0162, respectively for PARSEC
  \citep{parsec}, Victoria-Regina \citep{vandenberg2006},
  MIST \citep{mist0,mist1}, and YaPSI \citep{yapsi} isochrones.
\label{tomr}}
\end{figure}

\begin{figure}
\includegraphics[scale=0.8]{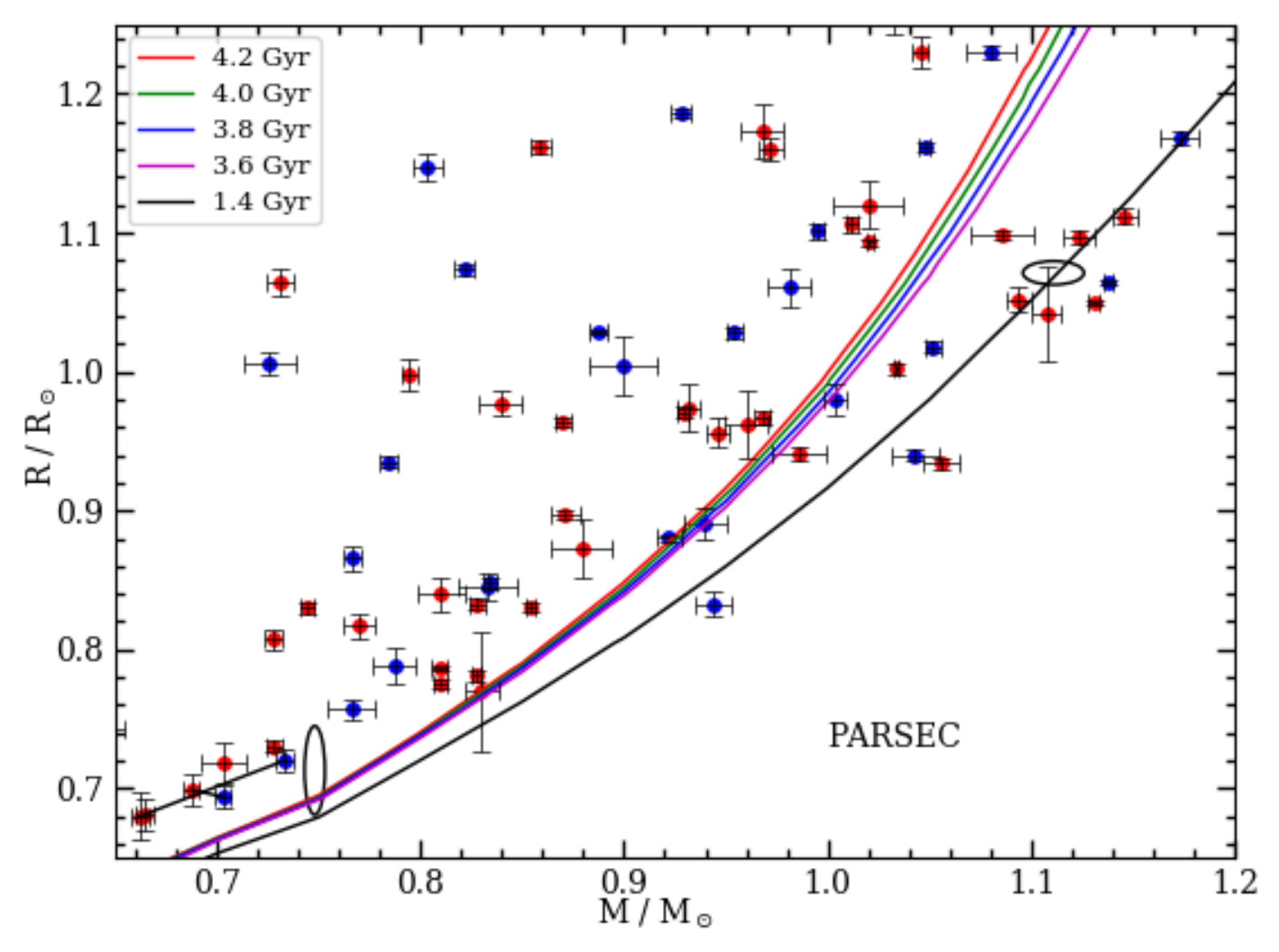}
\caption{Mass-radius plot for the members of S1247, along with
  precisely measured stars from eclipsing binaries
  \citep{debcat}. Isochrones have the same meaning as in
  Fig. \ref{tomr}. Primary stars in the eclipsing binaries are blue,
  and secondary stars are red. All error bars and ellipses use
  $2\sigma$ uncertainties. Black lines connect the members of binaries
  V404 CMa \citep{v404cma} and ASAS J082552-1622.8 \citep{asasbin},
  having stars with masses similar to that of WOCS 12009 B.
  \label{debs}}
\end{figure}

In Fig. \ref{tomr} we compare the results of our binary star model
fits with theoretical expectations from isochrones, and in
Fig. \ref{debs} we compare against precisely measured members of
eclipsing binaries \citep{debcat}\footnote{{\tt
    www.astro.keele.ac.uk/jkt/debcat/}, downloaded June 14, 2017}. For the comparisons in the
figures, the error ellipses for WOCS 12009 include contributions from statistical and
systematic error added in quadrature, and we have employed the chemical compositions used
by the modellers in calibrating their isochrones to the Sun.  The
properties of the secondary star (WOCS 12009 B) are in rough
agreement with isochrone predictions, although there is some
dependence on the chemical composition and the implementation of the
stellar physics in the models. Also of note, the secondary star is
slightly closer to the isochrones than any other star in the range from about
$0.7 - 0.8\msun$. This is probably because WOCS 12009 B is in a much
longer period binary than other measured main sequence stars. For
example, V404 CMa \citep{v404cma} has a period of just 0.452 d, and
ASAS J082552-1622.8 \citep{asasbin} has a period of 1.528 d. Tidal
interactions and the resulting stellar activity are likely to be
responsible for the relatively large radii in systems like those (see
\citealt{torres} and references therein). WOCS 12009 B is probably a
better approximation to a star that has evolved in isolation, and is
therefore valuable in its own right.

On the other hand, while the primary star (WOCS 12009 A) has
characteristics putting it among other well-measured eclipsing stars
in the field, it is much smaller than predicted for its mass and
reasonable ages for M67. Isochrone fits to the CMD have returned ages
from about 3.5 Gyr to 4.0 Gyr \citep{saraj} and 3.6 to 4.6 Gyr
\citep{vsm67}, with a large part of these ranges resulting from model
physics and chemical composition uncertainties. \citet{stellom67k2}
quote an age of about 3.5 Gyr from their asteroseismic analysis of pulsating
giants in M67.  The disagreement between WOCS 12009 A's radius and the
predictions from models with ages at the low end of the range
quoted for M67 is more than 15\%. The star's radius is larger
than that of a zero-age main sequence star, and is consistent with a
star of approximately 1.4 Gyr age.

\begin{figure}
\plotone{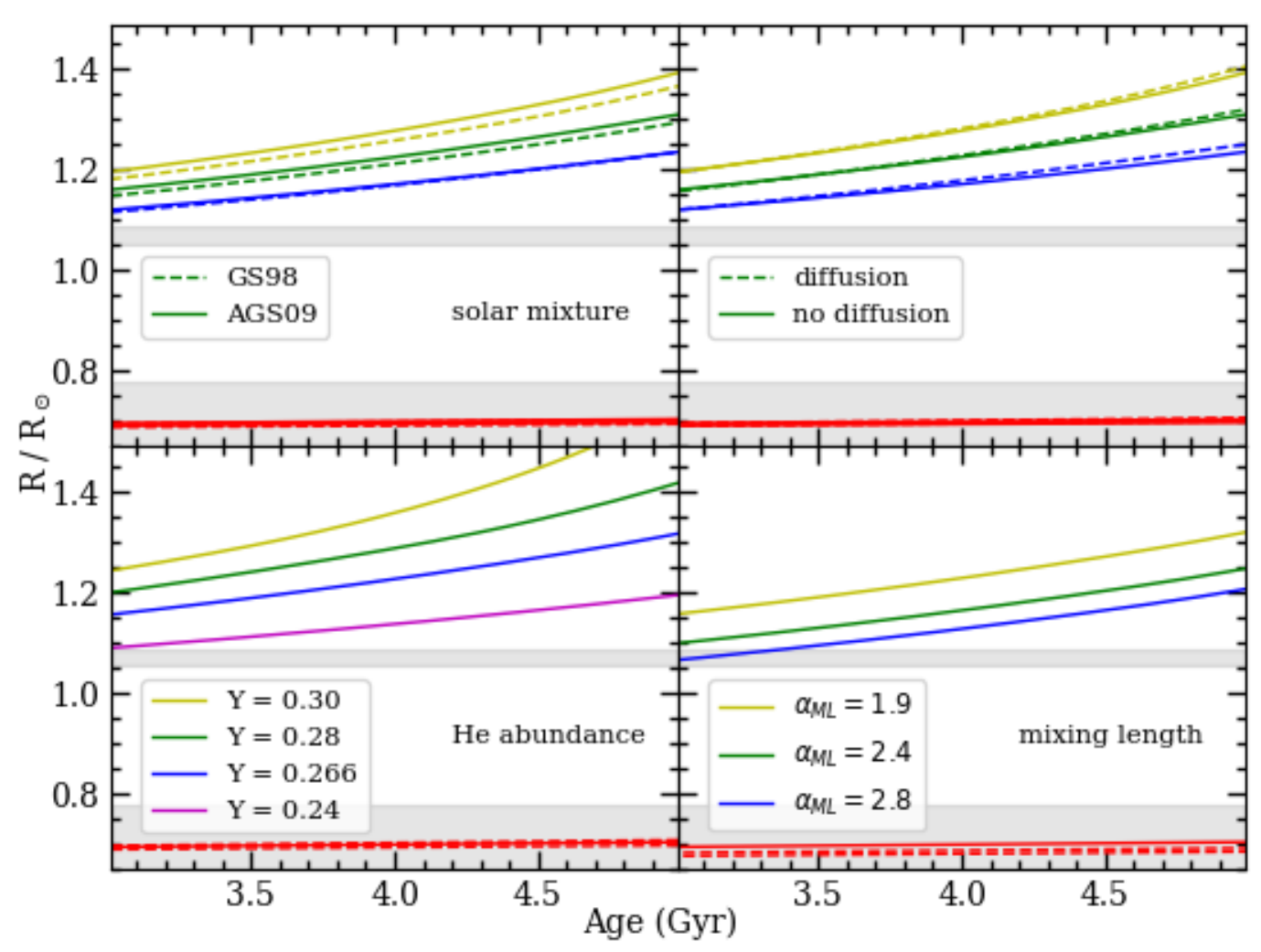}
\caption{Radius evolution of stars of $0.75\msun$ (red lines) and $1.12\msun$
  stars when model composition and physics inputs are varied.  Grey
  rectangles show the $2\sigma$ uncertainty range for the measured
  radii of WOCS 12009 stars. In the top two panels, tracks for [Fe/H]
  $=-0.1$ (yellow lines), 0.0 (green lines), and $+0.1$ (blue lines)
  are shown.\label{rmodel}}
\end{figure}

To illustrate the difficulties in explaining the mass and radius of
this star consistently with the cluster age, we explored
one-dimensional stellar evolution models (Fig. \ref{rmodel}).  Stellar
models were calculated using the 1-D Yale Rotating Evolution Code
(YREC; \citealt{yrec}), using the nuclear cross-sections, opacities,
equations of state, and model atmospheres described in section 2.3 of
\citet{somers}. For the base model, we adopt the \citet{asplund}
present-day solar photospheric abundance ratio $Z / X = 0.0181$, assume helium
and heavy metal diffusion as calculated by \citet{thoul}, calibrate
the solar mixing length and helium abundance to reproduce the solar
luminosity and radius at 4.57 Gyr, and assume no rotation. As
parameter variations, we test the alternate solar mixture given by
\citet{gs98}, assume diffusion-less envelopes, adopt different
proto-solar helium abundances, vary the convective efficiency by
changing the mixing length (e.g. \citealt{chabrier}), and permit
modest core overshooting of $\sim0.2$ pressure scale heights.

The secondary star's characteristics are less
sensitive to uncertainties in age, stellar physics, and composition
than the primary star's are, so that the overall agreement of models
and observations of the secondary is a useful validation. Reasonable
changes to the chemical composition of the stars are unable to explain
a radius discrepancy of this magnitude, although higher metallicity or
lower helium abundance would push the models in the right
direction. In the case of helium alone, the abundance would have to be
moved in a direction opposite what is observed for galactic chemical
evolution and reduced below Big Bang nucleosynthesis values ($Y =
0.245$) to come close to the observed radius. The metallicity of M67
stars has been repeatedly shown to be very close to solar, and so to
explain the radius discrepancy with metallicity would require a change
several times any reasonable uncertainty in M67's value.

Stellar physics uncertainties also seem to fall short of explaining
the small radius. The inclusion of diffusion makes a model star
marginally larger, and so this cannot alleviate the issue. More
efficient convection (through a larger mixing length parameter
$\alpha_{ML}$, for example) can produce a smaller star, but again, an
explanation of the radius issue would require a large change. For a
star with a thinner surface convection zone than the Sun, convection
is expected to be less efficient due to radiative losses. Convective
core overshoot tends to make the star slightly smaller, but even for
overshoot several times larger than inferred from cluster comparisons
(0.2 pressure scale heights), this effect is also too small by a
factor of at least three.

To summarize, we find no chemical composition or stellar physics
uncertainties that can explain the small size of the star WOCS 12009
A.

\subsection{Distance Modulus}\label{dm}

We can further test the membership of the binary by deriving the
luminosity of each star, and by comparing the absolute with the
apparent magnitudes.  The photometry of the two stars was deconvolved
using luminosity ratios from the multi-filter light curve analysis,
and photometric temperatures were derived from the colors and cluster
reddening (see section \ref{constrain}). The temperatures, along with
radii derived from the binary analysis and theoretical bolometric
corrections from \citet{cv}, allow us to derive the absolute
magnitudes $M_V$. We find $(m-M)_V = 9.74\pm0.03$ and $9.82\pm0.09$
for the primary and secondary stars respectively. The quoted
uncertainties include contributions from the measurement of the
apparent magnitudes, the calculation of the luminosity, and the
calculation of the bolometric correction. These values are in
excellent agreement with each other and with previous measurements of
M67's distance modulus. For comparison, some notable recent
measurements are $9.76\pm0.06\pm0.05$ (statistical and systematic
uncertainties) from 10 solar analogs \citep{pasquini}, $9.72\pm0.05$
using metal-rich field dwarfs with {\it Hipparcos} parallaxes
\citep{mephot}, and $9.70\pm0.04$ from K2 asteroseismology of more than 30 cluster
red giants \citep{stellom67k2}.

\subsection{The Color-Magnitude Diagram (CMD)}\label{cmd}

With well-studied binary stars, we can bring important and
rarely available new mass information to the study of the
color-magnitude diagram and to the comparison of observational data
with theoretical models.  To that end, we used luminosity ratios
derived from the binary star modeling to deconvolve the light in
different wavelength bands for the two WOCS 12009 stars, and determine
their CMD positions. In the discussions below, we use the
high-precision photometry of \citet{mephot} along with the
identifications for high-probability single-star members from that
study. The original list was selected based on being at the blue edge
of the band of main sequence and binary stars in the CMD, but we have
verified that the majority of the list were classified as single
members with no detectable radial velocity variability by
\citet{geller}. Those that were identified as binaries were
eliminated. So to the greatest degree we can, these stars should trace
out where the main sequence and turnoff should be for the age of M67.

\begin{figure}
\includegraphics[scale=0.8]{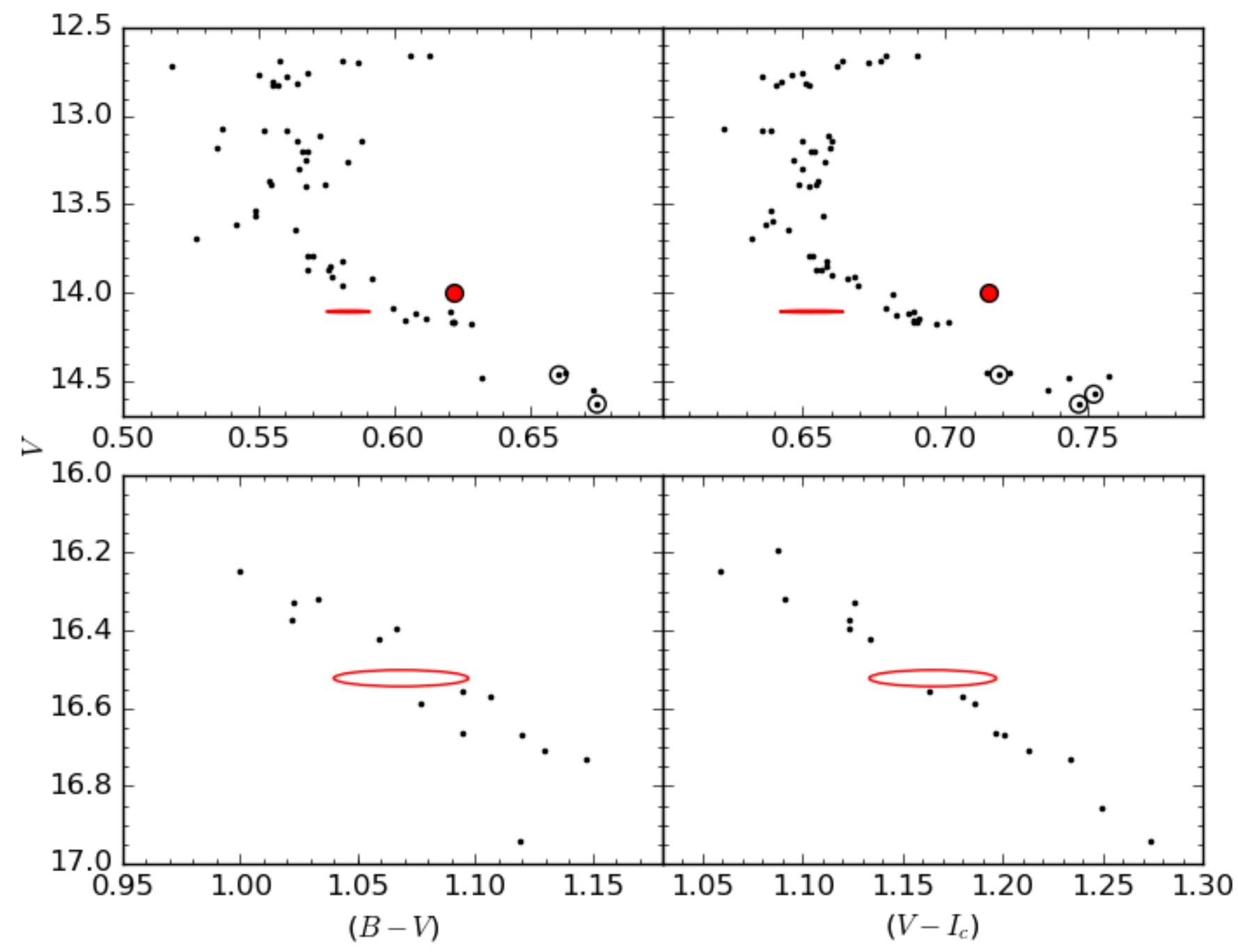}
\caption{Color-magnitude diagrams of M67 zoomed near the components of WOCS 12009.
Small points are probable single-star members with photometry from
  \citet{mephot} and circles are solar analogs identified by
  \citet{pasquini}. The system photometry for WOCS 12009 is shown with
  a red circle, and $2\sigma$ error ellipses for the deconvolved photometry of the component stars are
  shown in red. \label{zoomcomp}}
\end{figure}

Because the statistical uncertainties in the system photometry are
well below 0.01 mag and the system photometry has been calibrated in
the same way as the rest of the cluster stars we use in the CMDs, the
{\it relative} photometry should be
reliable.
In the
CMDs (see Fig. \ref{zoomcomp}), the secondary star falls among
probable single main sequence stars at $V \approx 16.5$. The primary
star is slightly bluer than the main sequence at $V \approx 14.1$, as
we would expect for a star that is less evolved than the other cluster
stars.

\begin{figure}
\includegraphics[scale=0.8]{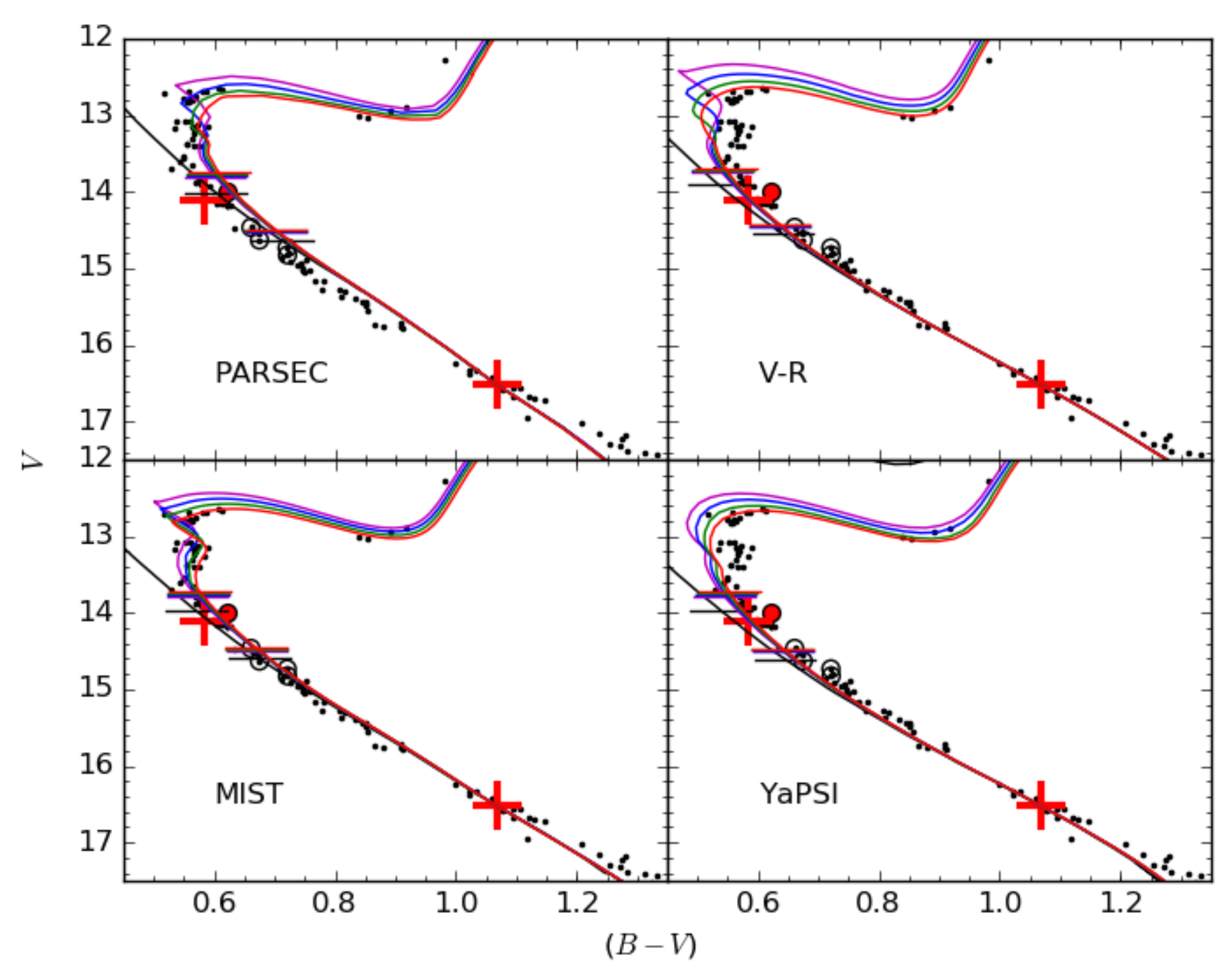}
\caption{($B-V$,$V$) color-magnitude diagram for the turnoff and main
sequence of M67. Small points are probable single-star members from
\citet{mephot} and circles are solar analogs identified by
\citet{pasquini}. The system photometry for WOCS 12009 is shown with
a red circle, and deconvolved photometry of the component stars are
red crosses.
Isochrones are from the PARSEC ($Z = 0.0152$),
Victoria-Regina ($Z = 0.0188$), MIST ($Z = 0.0142$), and
Yale-Potsdam ($Z = 0.01619$)
sets for ages of 1.4, 3.6, 3.8, 4.0, and 4.2 Gyr, and have been
shifted to match the mass and photometry of the secondary star of WOCS
12009. The theoretical predictions for stars with masses equal to the
primary star of WOCS 12009 ($V \approx 13.8 - 14.0$) and to the Sun
($V \approx 14.6 - 14.7$) are shown with horizontal
lines. \label{deconvolve}}
\end{figure}

\begin{figure}
\includegraphics[scale=0.8]{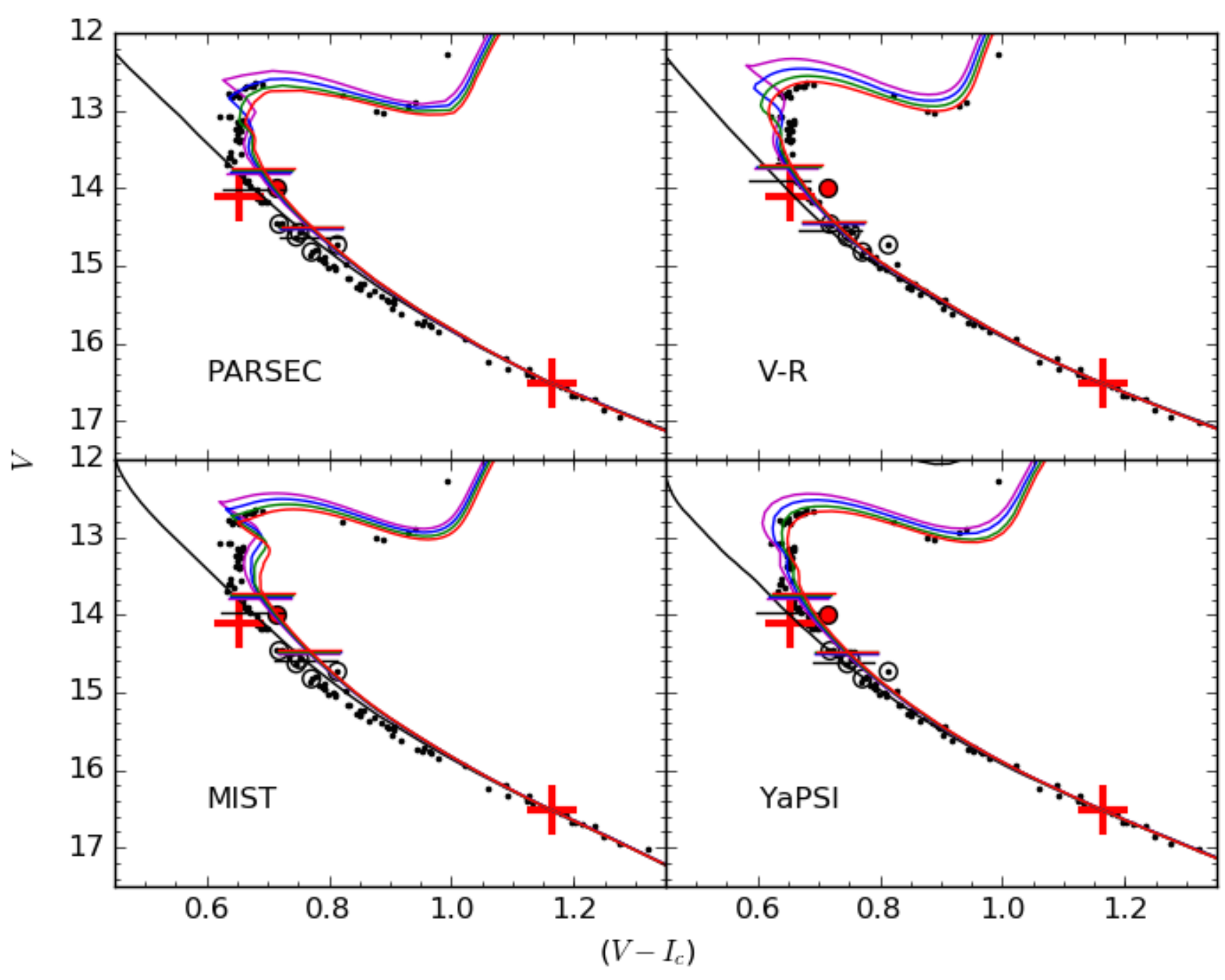}
\caption{Same as for Fig. \ref{deconvolve} except for ($V-I_C$,$V$) photometry.
\label{deconvolvevi}}
\end{figure}

When we shift isochrones to force a fit to the mass and photometry of
the secondary star (see Figures \ref{deconvolve} and
\ref{deconvolvevi}), there is decent agreement between the cluster
main sequence stars and the theoretical predictions for the V-R and
MIST isochrone sets.
The landscape is similar for the ($V-I_C$,$V$) CMD (see
Fig. \ref{deconvolvevi}), although the MIST isochrones don't match the main sequence as well.

The mass of the primary star would be a more valuable constraint on
the masses of evolved stars in the cluster and on the cluster's age if
it had had a conventional evolution history. However, if we utilize
the model isochrones in a relative sense, we can infer with some
confidence that a single cluster star with the mass of WOCS 12009 A
(about $1.11 \msun$) and an uneventful evolution history should reside
at $V \approx 13.8$, about 0.35 mag brighter than WOCS 12009 A and
just faintward of the ``hook'' feature in the CMD. Models tied to WOCS
12009 B also predict that a solar mass star in M67 should have $V
\approx 14.57$. This is quite close to the $V$ magnitude of the star
Sanders 1095 (1787 in \citealt{yadav}) that \citet{pasquini}
identified as one of their two best solar analogs.

In the CMD, the turnoff of M67 potentially gives us a lot of insight
into the physics of the stellar cores. Our group is currently working
on additional binary systems that will allow us to identify the
precise masses of stars at the turnoff, and so we will not undertake a
full discussion of the topic here. The CMDs in Figures
\ref{deconvolve} and \ref{deconvolvevi} illustrate the substantial
differences in the isochrones produced by different groups, and
without clarity on the ``correct'' physics and chemical composition
inputs to include in the models, systematic errors in the cluster age
will be unavoidable.

The noticeable gap seen at $V \approx 12.9$ is an indicator of rapid
evolution through that part of the CMD, probably due to the presence
of a small convective core maintained by CNO cycle hydrogen burning
reactions and their strong temperature dependence. However, there are
several composition and physics factors that can affect the size (and
even the presence) of a convective core in models. \citet{vg07} and
\citet{magic} studied extensive sets of stellar models to try to
identify the important factors, but \citeauthor{magic} in particular
found that there are degeneracies in how the CMD is affected.
Composition factors like the metallicity of M67 relative to the Sun
and the actual solar mixture affect the abundance of CNO cycle
catalyst nuclei, while physics like diffusion, the prescription for
convective core overshoot, and CNO cycle reaction rates can affect the
strength of the energy release in the stellar cores. While some of the
models approximately reproduce the brightness of the gap (which occurs
in the straight-line segment immediately faintward of the bluest point
on the isochrone), models generally are not as successful in matching
the blue protrusion among the M67 main sequence stars at $V \approx
13.7$. To the extent that we are able to check with current models,
the cluster CMDs are consistent with typically quoted ages of
$3.8-4.2$ Gyr.

\subsection{Stellar Motions and Li Abundance}

Every indication we have is that the binary WOCS 12009 is a member of
the M67 cluster, but the primary star is significantly smaller than
expected for reasonable cluster ages, and it is also hotter and less
luminous than expected for its mass. Before discussing possible
explanations, we assemble other observations that can comment on the puzzle.

Using our highest resolution spectroscopic observations (with the TRES
and HARPS-N spectrographs), we looked at the rotational motions of the
two stars in the binary, but were unable to detect rotation in the
broadening functions of the two stars. The upper limits are $v_{\rm rot}
\sin i < 5$ \kms from the TRES observations and $< 4$ \kms from the
HARPS-N observations. Although slow rotation is expected for
single solar-type stars in an old cluster like M67, the low rotation speed of
the primary has some impact on our understanding of its genesis, as
discussed later.

\begin{figure}
\includegraphics[scale=0.7]{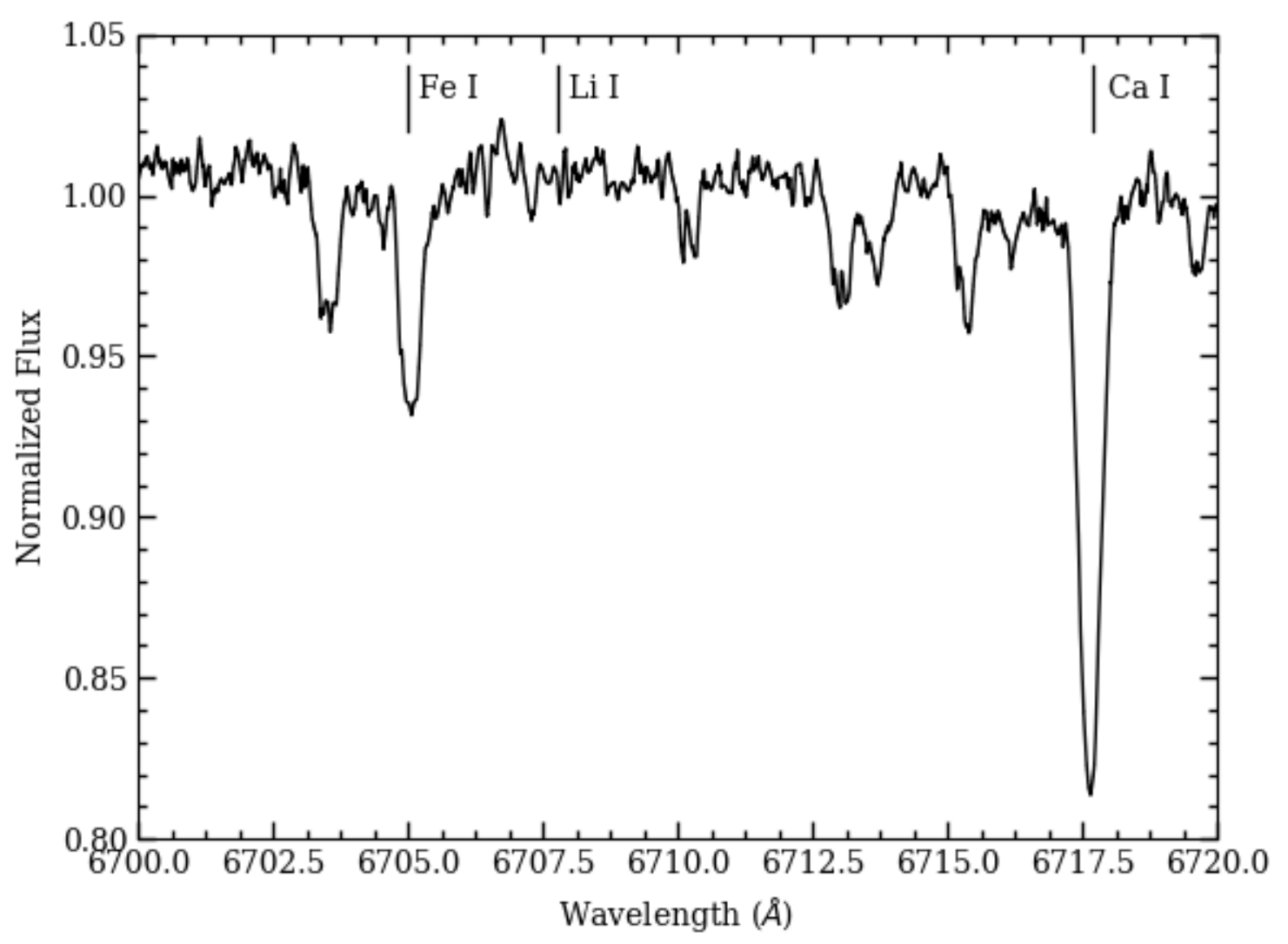}
\caption{The averaged FLAMES spectrum for WOCS 12009 A in the vicinity of 
  the Li I resonance doublet. \label{lispec}}
\end{figure}

\begin{figure}
\includegraphics[scale=0.8]{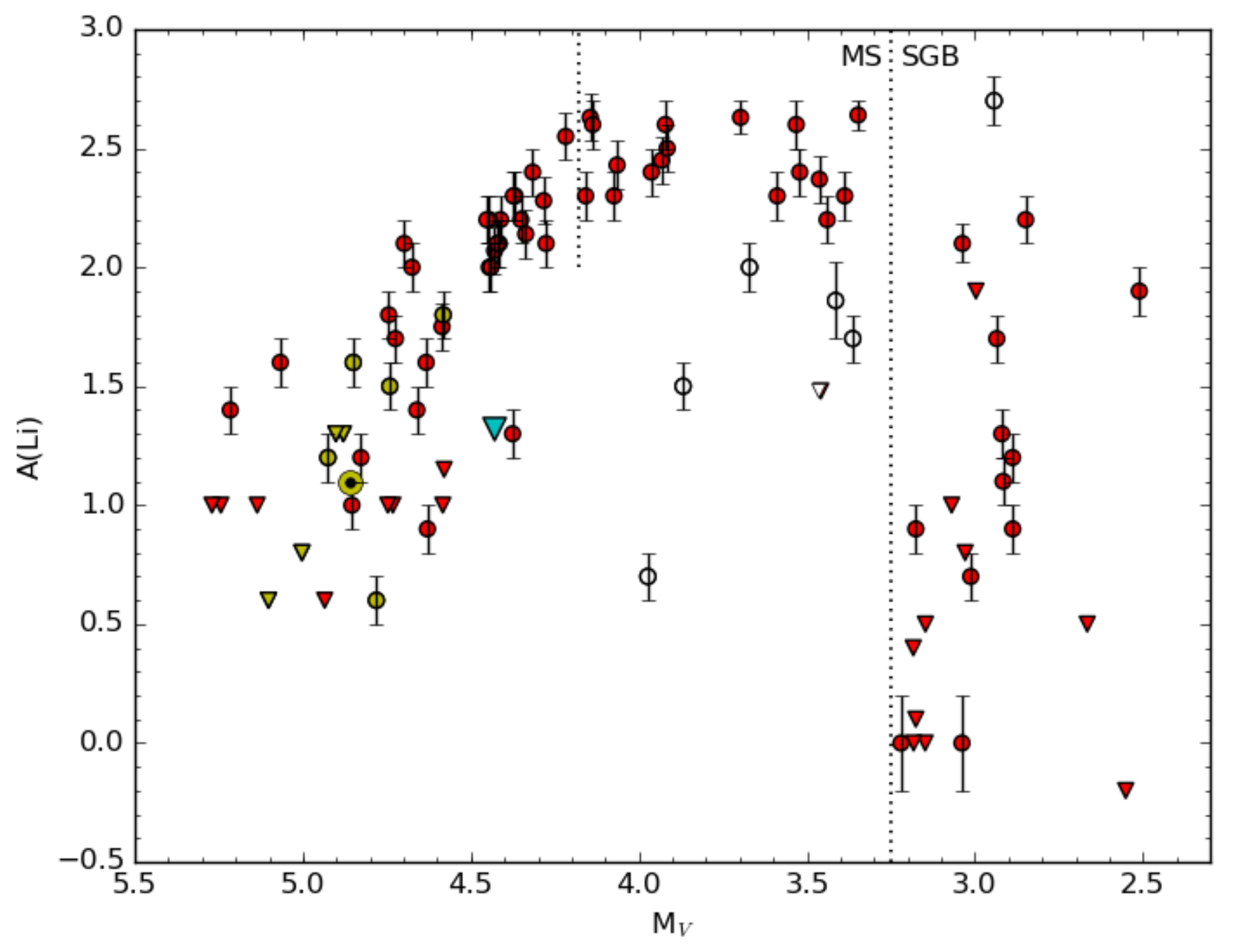}
\caption{Li abundances ($A(\mbox{Li}) = \log (N_{\mbox{Li}} /
  N_{\mbox{H}}) + 12$; \citealt{pace}) as a function of absolute
  magnitude $M_V$, assuming a distance modulus $(m-M)_V = 9.72$. Red
  symbols indicate probable single stars, yellow symbols are the Sun
  ($\odot$) and solar twin candidates \citep{pasquini}, red
  triangles are upper limits, and white symbols are stars identified
  as deviant by \citeauthor{pace}. The blue arrowhead is the upper
  limit for WOCS 12009 A \citep{jones}. The dotted line at $M_V =
  4.19$ shows the expected position of normal cluster star with WOCS
  12009 A's mass. \label{Liplot}}
\end{figure}

The mass of WOCS 12009 A should place it at the faint end of the
cluster's Li plateau with abundances $A(\mbox{Li}) \approx 2.5$
\citep{pace}. Stars in the plateau have outer convective zones that do
not mix Li atoms down to temperatures where nuclear reactions can
burn them, and additional mixing processes must have little if any
effect. Even with the diluting effects of the secondary star's light
on the Li lines, Li should have been detectable if it was present with
the plateau abundance. Our spectral disentangling allows us to mostly
remove the secondary star's contribution to the spectrum and to
produce an average spectrum for the primary star as a byproduct. The
averaged FLAMES spectrum for the primary is shown in
Fig. \ref{lispec}, as it has the highest signal-to-noise in that
wavelength range. There is still no sign of the Li lines for the
primary star. \citet{jones} provide an upper limit for the system
($A(\mbox{Li}) < 1.32$) that puts the star's abundance at least about
0.8 dex lower than stars at the same absolute magnitude, but because
the star is probably underluminous compared to normal cluster stars of
the same mass (see \S \ref{cmd}), the discrepancy is more than an
order of magnitude (Fig. \ref{Liplot}).

\subsection{The Formation of the Primary Star}

The Li observations could be explained by a
stellar union: stars of about $0.95 \msun$ or less would have
depleted their surface Li to undetectability well before the present
age of the cluster because the surface convection zone easily reaches
down to Li burning temperatures.\footnote{Be burns at slightly higher
  temperatures, and observations of the Be abundance could corroborate
  this result. However, the most commonly used lines at 313.04 and
  313.11 nm do not fall in the observed wavelength range for any of
  our spectra.}  In the union of two such stars, a more massive star
with depleted Li would result.

This could also explain the smaller than expected radius of the
primary star. In hydrodynamic simulations of stellar collisions, it is
found that the material from the two stars gets sorted radially in the merger
remnant by entropy to good approximation \citep{lombardi}. Because
lower mass stars have lower temperatures and higher densities in their
cores, the core of the lower mass star ends up at the core of the
remnant. This effectively restarts the remnant with a higher core
hydrogen content, and makes it look like a younger main sequence star
once a Kelvin-Helmholtz timescale has passed and the remnant has
thermally adjusted to its new structure. As a result, if the star's
size makes it look like it is 1.4 Gyr old now, that is likely to be
fairly close to the time elapsed since a merger took place.

Two stars could combine as part of the merger of a close binary pair
or the collision of two stars during a resonant multi-star interaction.
Both are likely to produce a remnant that is depleted of surface Li,
but the route for combining the stars could leave signatures on their
motions. Is there a reasonably probable mechanism that could produce
a slowly rotating remnant star in wide low eccentricity ($e =
0.05$) binary?

\begin{figure}
\includegraphics[scale=0.8]{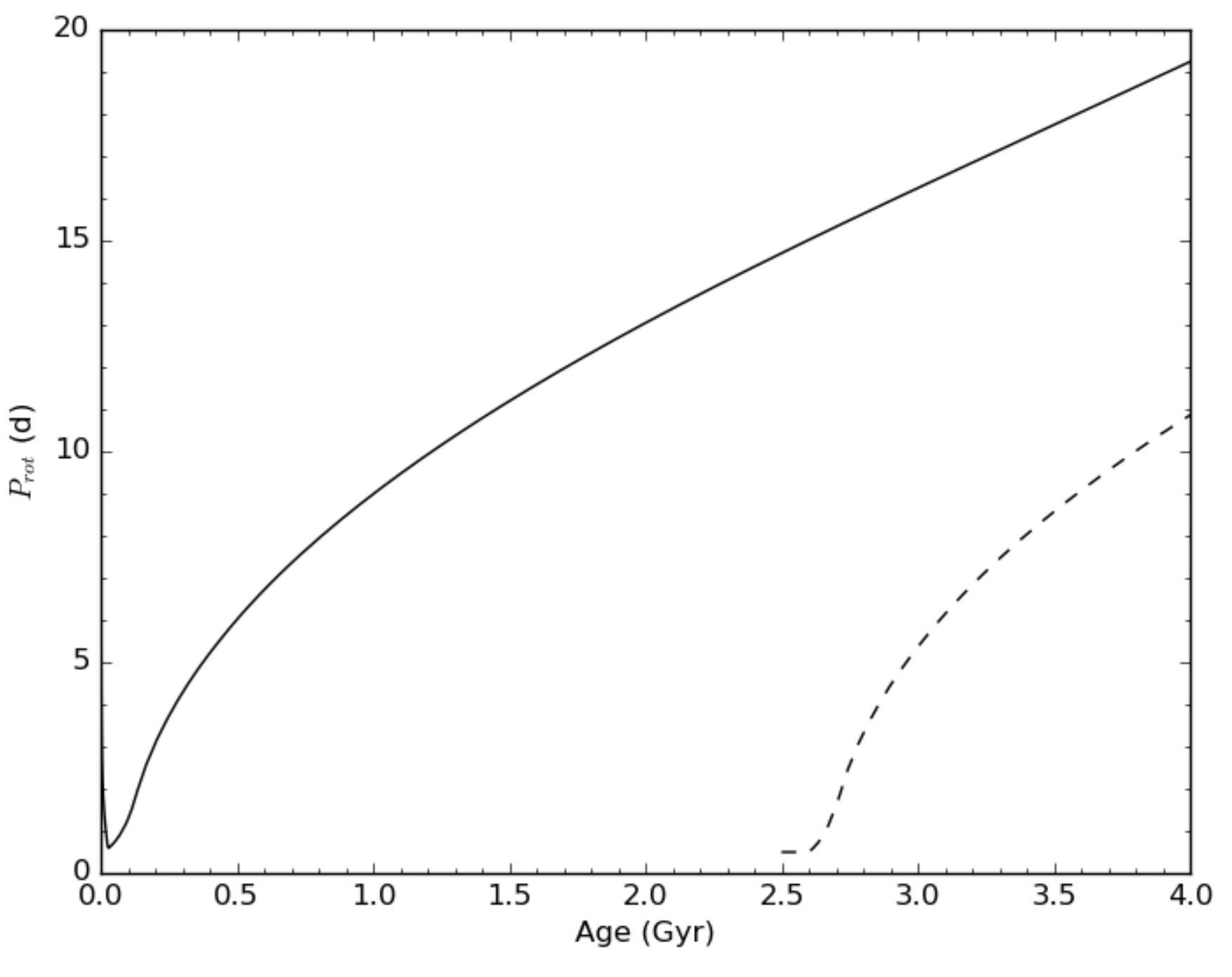}
\caption{Rotational evolution for $1.1 \msun$ stellar models having a
  single-star evolutionary history ({\it solid line}) and having a
  birth as a stellar merger 1.5 Gyr before the present day ({\it
    dashed line}). \label{rotmod}}
\end{figure}

Unless the combining stars were in a nearly head-on collision, the
remnant star is likely to have started with rapid rotation. The
primary star in WOCS 12009 probably has a surface convection zone
though, meaning that some spindown via a magnetic wind is likely. If
the remnant was made approximately 1 Gyr ago, we can ask whether there
was sufficient time for it to spin down as an effectively single
star. The rotation of stars in the 1 Gyr old cluster NGC 6811 was
studied during the \kep mission \citep{meibom11}. In that cluster,
stars with $(B-V)$ color similar to WOCS 12009 A have spun down from
their zero-age main sequence values to a period of around 10 d. For a
star the size of WOCS 12009 A, a 10 d period roughly corresponds to
rotation near our $v_{\rm rot} \sin i$ detection limit of 5 km
s$^{-1}$. We computed rotational stellar models using YREC (see
section \ref{mrsec}) to check this idea further, as shown in
Fig. \ref{rotmod}.  A $1.1 \msun$ star has enough of a surface
convection zone that spindown via a magnetic wind could bring a
rapidly rotating star (modeled here with a period $P_{\rm rot} = 0.5$
d) down to our detection limit in 1.5 Gyr.  A detected rotation period
for the primary, as opposed to our current upper limit, could in
principle be used to infer the time since merger, for the same reason
that a rotation period for a single non-interacting star can be used
as an age indicator. This would be one interesting test of the merger
hypothesis.

It is plausible that a rapidly rotating merger remnant could have spun
down to the levels we see today, but other observations may test
this. For example, the Ca II H and K emission index (an activity
indicator) has been measured for WOCS 12009 by \citet{giampapa}, and
they find that it has a relatively high value relative to other M67
stars, but not out of the range of the best vetted single star
candidates \citep{curtis}. \citet{mama} compute $\log R\arcmin_{\rm
  HK} = -4.70$ from the \citeauthor{giampapa} data, which probably
makes it more active than the 95\% range for the Sun.  An examination
of spot modulation on the primary star might also be able to reveal
whether it is rotating significantly more rapidly than single cluster
stars of similar color. Such measurements are currently underway under
the umbrella of the K2 M67 Study.

The other major question is whether the orbital characteristics of the
current binary could be natural products of any potential formation
mechanisms. Below we consider stellar collisions and the evolution of
hierarchical triple systems.

A stellar collision during a multi-body dynamical interaction is a
relatively common outcome if the incoming binary (or binaries) are
tightly bound.  \citet{fregeau} examined binary-single and
binary-binary scattering experiments with collisions allowed, and
found as expected that low eccentricities are strongly disfavored in
binaries formed after stellar collisions --- the conditions required for
the outgoing stars to wind up on a nearly circular orbit are too
narrow.  Even with a maximal amount of expansion by a merger remnant
before it thermally relaxes, tidal interactions are unlikely to be
able to help circularize the orbit of WOCS 12009, which is currently
relatively wide ($a / R_1 \approx 82$).

Formation involving a hierarchical triple system that dynamically
formed as part of a multi-body interaction along with the subsequent
merger of the close pair faces some of the same difficulties. For
guidance, we examined the scattering study of \citet{antognini}. Those
authors find that when triple systems are dynamically formed, a
compact configuration (with the ratio of the outer orbit periastron to
the inner orbit semi-major axis $r_{\rm P,outer} / a_{\rm inner} \sim
10$) is a preferred outcome. If the current orbit of WOCS 12009
corresponds to the outer orbit of such a compact triple, the inner
orbit could have had a period of around 2 d. This is long enough that
the inner binary would not have merged rapidly via tidal interactions
or magnetic braking.  However, an eccentricity as low as 0.05 for the
outer orbit is again strongly disfavored as part of the dynamical
formation of the triple.

A primordial triple system with an inner binary that merges after
about 3 Gyr is also a possibility. First considering the low
eccentricity of the current orbit, we note that low eccentricity
binaries with periods above $10^4$ d are known among solar-type
binaries in the field \citep{raghavan}.\footnote{For orbital periods
  less than 1000 d, solar-type stars in binaries appear to have a
  relatively flat distribution of eccentricities, but most of those
  with $e < 0.1$ have periods of less than 12 d and are likely to have
  been produced by tidal interactions.} A few percent of binaries with
orbital solutions in the 2.5 Gyr old cluster NGC 6819
\citep{milliman}, in M67 \citep{pollack}, and the 7 Gyr old cluster
NGC 188 \citep{gellern188} are known to have $e < 0.1$ and $P > 20$
d. (For NGC 6819 and M67, there are known long-period systems with $e
< 0.06$.) For known triple systems, the outer orbit is usually
eccentric, but there are known examples with $e < 0.05$. So it is not
out of the question to think that the low eccentricity can result from
the star formation process.

The inner binary might have formed with an initial period near 2 d as
part of the star formation process, allowing angular momentum loss
mechanisms like tides and magnetic braking \citep[][e.g.,]{stepien} to
produce a merger. The majority of the time ($\sim 2-3$ Gyr) would be
needed to have the binary lose enough energy to come into contact and
ultimately merge.  M67 systems AH Cnc, EV Cnc, and HS Cnc are contact
or near-contact binaries near the cluster turnoff
\citep{stassun,ss2003,pribulla,yakut} that could be in this stage of
their orbital evolution, and theoretical studies indicate that
sub-turnoff mergers could be of order 3-4\% of the upper main sequence
population for an old cluster like M67 \citep{andronov}.

A primordial triple with an inner binary having a period substantially
longer than 3 days might also be capable of producing a system like
WOCS 12009, but stability against disruption is a concern if the inner
binary has too long a period. As a guide, \citet{mardling} used a
criterion for stability against escape of a component star that
implies $r_{\rm P,outer} / a_{\rm inner} \sim 3.4$ for the WOCS 12009
system, which translates to an inner orbit with $a_{\rm inner} \approx
0.12$ AU or $P_{\rm inner} \approx 14$ d.  For an inner binary formed
with an initial period closer to 14 d, Kozai-Lidov oscillations due to
the third star would probably be needed to drive the binary's
eccentricity up before tidal dissipation could produce a shorter
period \citep{fabry}. This would require misalignment of the inner and
outer orbital planes, but Kozai-Lidov cycles can have a very short
timescale (comparable to but longer than the period of the outer
orbit), so that there would be no difficulty with producing a close
binary within the age of M67.  Once close binary interactions become
dominant, Kozai cycles are suppressed and the binary is expected to
evolve as if it is isolated from the third star.

For a stable primordial triple formed with an outer orbital period
near 70 d, either of these paths would likely result in a circularized
inner binary with a period of a few days, which will later merge.
If a Kozai mechanism was responsible for the initial stage of
decreasing the separation of an inner binary system, we might be able
to observe the signature of an inclination difference between the
orbit of the inner binary and the orbit of the teriary star
\citep{fabry}. The rotation axis of the merged star would probably
retain the direction of the angular momentum vector for the close
binary. Because the current system reflects the orbital plane of the
outer binary in the original triple and because the binary is
currently eclipsing, the rotation axis of the primary star could be
inclined to our line of sight by about $40\degr$. Thus, depending on
its axial orientation, it might be possible to observe muted
rotational effects such as the Rossiter-McLaughlin effect on the
radial velocities or rotational modulation of the light curve. 

\section{Conclusions}

We present high-precision measurements of the masses, radii, and
photometry for stars in the bright detached eclipsing binary system
WOCS 12009 in the cluster M67. This cluster is commonly used as a
testbed for stellar evolution models of solar metallicity, and the
addition of new observables for stars in the cluster will further test
the fidelity of the physics we use to model them. We are currently in
the process of analyzing other eclipsing cluster binaries that will
expand these tests to higher and lower mass stars in M67.

For the WOCS 12009 system, we find that the primary star is
substantially smaller than expected from any reasonable evolution
model for a single star of its mass. This, as well as the
non-detection of photospheric Li, and its relatively high temperature
and low luminosity, points us toward an explanation involving a merger
of two lower mass stars. The low rotation rate of the primary star
appears to be explicable if a merger took place more than about a
billion years ago, as this would give the remnant star time to spin
down as an effectively single star. The long-period, low eccentricity
orbit of the binary is harder to understand, as most dynamical
interactions in the cluster should have resulted in eccentricity that
could not have been eliminated in the age of the cluster. Our
preferred explanation is that this was born as a compact hierarchical
triple system, and tidal interactions (possibly assisted by
Kozai-Lidov oscillations) eventually brought the system into contact
and led to a merger.

Regardless of the mechanism behind its formation, the primary star
might have been identified as a blue straggler had it not had a cool
companion that reddened the binary's combined light. Even so, the
primary star has a low enough mass that it is only slightly bluer than
the cluster main sequence. However, as the cluster ages and single
stars of the same mass evolve off the main sequence, this star will
retain its youthful main sequence appearance for an extra $2-3$
billion years and will be more clearly identifiable as a product of
unusual circumstances.

\acknowledgments E.L.S. is grateful to K. Brogaard for providing the
original version of the spectral disentangling code used in this
work. D.S. is the recipient of an Australian Research Council Future
Fellowship (project number FT1400147). G.S. acknowledges the support
of the Vanderbilt Office of the Provost through the Vanderbilt
Initiative in Data-Intensive Astrophysics (VIDA)
fellowship.  M.H.P. acknowledges support from NASA ADP grant
NNX15AF13G. We would also like to thank J. Valenti, D. Soderblom, and
S. Basu for helpful conversations during the course of the project.

This paper includes data collected by the {\it K2}
mission, and we gratefully acknowledge support from NASA under grant
NNX15AW69G to R.D.M. Funding for the {\it K2} mission is provided by
the NASA Science Mission Directorate.

This research made use of observations from the Las Cumbres
Observatory network; the SIMBAD database, operated at CDS, Strasbourg,
France;
the WEBDA database, operated at the
Institute for Astronomy of the University of Vienna;
the ESO Science Archive Facility under request number 289799;
and the Mikulski Archive for Space Telescopes (MAST). STScI is operated by the
Association of Universities for Research in Astronomy, Inc., under
NASA contract NAS5-26555. Support for MAST is provided by the NASA
Office of Space Science via grant NNX09AF08G and by other grants and
contracts.

Funding for SDSS-III has been provided by the Alfred P. Sloan
Foundation, the Participating Institutions, the National Science
Foundation, and the U.S. Department of Energy Office of Science. The
SDSS-III web site is http://www.sdss3.org/.

SDSS-III is managed by the Astrophysical Research Consortium for the
Participating Institutions of the SDSS-III Collaboration including the
University of Arizona, the Brazilian Participation Group, Brookhaven
National Laboratory, Carnegie Mellon University, University of
Florida, the French Participation Group, the German Participation
Group, Harvard University, the Instituto de Astrofisica de Canarias,
the Michigan State/Notre Dame/JINA Participation Group, Johns Hopkins
University, Lawrence Berkeley National Laboratory, Max Planck
Institute for Astrophysics, Max Planck Institute for Extraterrestrial
Physics, New Mexico State University, New York University, Ohio State
University, Pennsylvania State University, University of Portsmouth,
Princeton University, the Spanish Participation Group, University of
Tokyo, University of Utah, Vanderbilt University, University of
Virginia, University of Washington, and Yale University.

\facilities{FLWO:1.5m (TRES), WIYN (Hydra), TNG (HARPS-N), LCOGT, MLO:1m}

\software{K2SC pipeline \citep{k2ox}, IRAF \citep{iraf1,iraf2},
  HARPS Data Reduction Software (v3.7), ELC \citep{elc}, (YREC;
  \citealt{yrec})}

\begin{deluxetable}{lcccl}
\tablecolumns{4}
\tablewidth{0pc}
\tablecaption{Ground-Based Photometric Eclipse Observations}
\tablehead{
\colhead{UT Date} & \colhead{Filters} & \colhead{mJD\tablenotemark{a} Start} &
\colhead{$N_{obs}$} & \colhead{Telescopes\tablenotemark{b}}}
\startdata
2003 Apr. 21 & $I_c$ & 2750.63952 & 150 & MLO \\
2004 Jan. 26 & $V$ & 3029.63023 & 168 & LaS \\%
2016 Jan. 1 & $BVI_c$ & 7389.40312 & 58,61,62 & CPT \\
2016 Feb. 3 & $BVR_c$ & 7422.55368 & 98,146,96 & ELP,LSC,MLO \\
2016 Apr. 13 & $BVI_c$ & 7492.22555 & 72,69,77 & CPT \\ 2016 May 20 &
$BVI_c$ & 7528.65024 & 30,31,33 & MLO,OGG\\
\enddata
\tablenotetext{a}{mJD = BJD - 2450000}
\tablenotetext{b}{MLO: Mount Laguna Observatory 1m. LaS: La Silla
  Observatory 1.5m. {\bf LCOGT Facilities:} CPT: South Africa
  Astronomical Observatory \#10, 13. ELP: McDonald Observatory 1m
  \#8. LSC: Cerro Tololo Inter-American Observatory \#4, 5, 9. OGG:
  Haleakala Observatory 0.4m \#6.}
\label{tableobs}
\end{deluxetable}

\begin{deluxetable}{rcccc|rcccc}
\tablewidth{0pt}
\tabletypesize{\scriptsize}
\tablecaption{Radial Velocity Measurements}
\tablehead{\colhead{mJD\tablenotemark{a}} & \colhead{$v_A$} & \colhead{$\sigma_{A}$} & \colhead{$v_B$} & \colhead{$\sigma_B$} & \colhead{mJD\tablenotemark{a}} & \colhead{$v_A$} & \colhead{$\sigma_{A}$} & \colhead{$v_B$} & \colhead{$\sigma_B$}\\
& \multicolumn{2}{c}{(km s$^{-1}$)} &  \multicolumn{2}{c}{(km s$^{-1}$)} & &  \multicolumn{2}{c}{(km s$^{-1}$)} &  \multicolumn{2}{c}{(km s$^{-1}$)}}
\startdata
\multicolumn{5}{c}{CfA Observations} & \multicolumn{5}{c}{WIYN Observations}\\
47200.90558 & 51.31 & 0.66 & &            & 54100.95996 & 55.15 & 0.16 &  4.93 & 1.13 \\
47216.87938 & 15.03 & 0.66 & &            & 57407.00468 &  7.16 & 0.12 & 73.96 & 1.04 \\
47230.82288 & 12.04 & 0.66 & 68.78 & 2.26 & 57443.91188 & 58.43 & 0.26 & -3.08 & 1.19 \\
47489.89727 & 27.54 & 0.66 & &            & 57444.90238 & 58.17 & 0.21 & -1.83 & 1.62 \\
47493.89927 & 18.58 & 0.66 & &            & 57445.88308 & 57.07 & 0.19 & -0.97 & 1.00 \\
47495.93427 & 13.24 & 0.66 & 60.08 & 2.26 & 57446.87188 & 56.63 & 0.12 &  1.63 & 0.46 \\
47513.97706 & 17.29 & 0.66 & 52.15 & 2.26 & 57447.91718 & 55.21 & 0.13 &  2.90 & 0.80 \\
47524.02176 & 39.05 & 0.66 & 25.05 & 2.26 & 57448.89118 & 54.27 & 0.21 &  7.97 & 2.13 \\
47543.94386 & 57.58 & 0.66 & -3.66 & 2.26 & 57762.88603 & 16.68 & 0.16 & 58.84 & 0.41 \\
47555.88246 & 36.83 & 0.66 & &            & 57764.00282 & 18.38 & 0.18 & 56.75 & 0.63 \\
47579.82666 & 10.70 & 0.66 & 65.36 & 2.26 & \multicolumn{5}{c}{HARPS-N Observations}\\    
47609.78406 & 58.44 & 0.66 & -4.78 & 2.26 & 57752.73729 &  7.25 & 0.055 & 74.59 & 2.01 \\
47633.71796 & 17.82 & 0.66 & 55.80 & 2.26 & 57780.44623 & 50.45 & 0.027 &  9.46 & 0.56 \\
47636.74956 & 12.20 & 0.66 & 65.80 & 2.26 & \multicolumn{5}{c}{APOGEE Observations}\\     
47846.96906 &  9.36 & 0.66 & 64.30 & 2.26 & 56654.92723 & 31.01 & 0.19 & 37.24 & 0.83 \\  
47898.98907 & 49.28 & 0.66 & &            & 56668.86973 & 55.17 & 0.14 & -0.90 & 0.73 \\  
47928.94617 & 10.47 & 0.66 & &            & 56698.75512 & 17.31 & 0.24 & 57.44 & 1.65 \\  
47994.75116 &  7.23 & 0.66 & &            & 56672.86367 & 58.64 & 0.64 & -2.56 & 0.64 \\  
48023.69056 & 55.46 & 0.66 & &            & 56677.88291 & 57.16 & 0.07 & -2.51 & 0.68 \\  
48024.64486 & 58.09 & 0.66 & &            & 56700.76763 & 12.64 & 0.12 & 65.24 & 1.14 \\  
48281.93347 & 19.56 & 0.66 & &            & 56734.66629 & 50.21 & 0.33 &  8.88 & 0.6 \\   
48291.90987 & 39.26 & 0.66 & &            & 56762.63516 & 31.02 & 0.20 & 36.42 & 0.6 \\   
54424.00437 & 30.99 & 0.66 & &            & \multicolumn{5}{c}{FLAMES Observations}\\    
\multicolumn{5}{c}{TRES Observations}     & 54137.78614 & 17.07 & 0.13 & 57.97 & 0.35\\  
57362.92518 & 51.81 & 0.037 &  7.97 & 0.41 & 54142.68654 & 27.07 & 0.44 & 44.55 & 0.52\\  
57403.90238 &  7.48 & 0.040 & 71.57 & 2.88 & 54154.58411 & 49.96 & 0.17 &  9.13 & 0.23\\  
57405.85138 &  7.34 & 0.043 & 73.85 & 0.36 & \multicolumn{5}{c}{Keck HIRES Measurement}\\ 
57406.92938 &  7.63 & 0.047 & 72.88 & 0.67 & 50143.94133 & 17.4  & & & \\                 
57407.90858 &  8.11 & 0.047 & 71.78 & 0.38 \\
57408.81588 &  8.80 & 0.026 & 71.77 & 0.46 \\
57436.77248 & 56.43 & 0.035 & -0.57 & 0.43 \\
57450.68408 & 51.18 & 0.047 &  8.10 & 0.55 \\
57714.99979 & 55.83 & 0.038 &  1.00 & 0.64 \\
57746.91989 & 13.08 & 0.056 & 64.48 & 0.72 \\
\enddata
\label{spectab}
\tablenotetext{a}{mJD = BJD - 2400000.}  
\end{deluxetable}

\begin{deluxetable}{lccclc}
\tablewidth{0pt}
\tablecaption{Out-of-Eclipse Photometry for WOCS 12009 \label{phottab}}
\tablehead{\colhead{Filter} & \colhead{S04} & \colhead{YBP} & \colhead{MMJ} & \colhead{Filter} & \colhead{2MASS}}
\startdata
$B$ & $14.6161\pm0.0038$ & $14.646\pm0.008$ & $14.663$ & $J$ & $12.785\pm0.021$ \\
$V$ & $13.9942\pm0.0005$ & $14.007\pm0.003$ & $14.044\pm0.05$ & $H$ & $12.459\pm0.023$ \\
$I_C$ & $13.2792\pm0.0007$ & $13.283\pm0.001$ & $13.266$ & $K_S$ & $12.422\pm0.023$ \\
\enddata
\tablecomments{S04: \citet{mephot}.  YBP: \citep{yadav}. MMJ: \citet{mmj}. 2MASS: \citet{2mass}.}
\end{deluxetable}

\begin{deluxetable}{lcccc}
\tablewidth{0pt}
\tabletypesize{\scriptsize}
\tablecaption{Best-Fit Model Parameters for WOCS 12009}
\label{parmtab}
\tablehead{\colhead{Parameter} & \multicolumn{3}{c}{K2 Light Curves/Limb Darkening Law} & \colhead{Combined}\\
& \colhead{PSF-based Approach LC} & \colhead{K2SC LC} & \colhead{K2SFF LC} & }
\startdata
$T_1$ (K) & \multicolumn{4}{c}{$6100\pm100$ (constraint)} \\%
\hline
$\gamma_1$ (\kms) & $33.815\pm0.009$ & $33.814\pm0.009$ & $33.815\pm0.009$ & $33.815\pm0.009$\\
$\gamma_2$ (\kms) & $33.745\pm0.010$ & $33.746\pm0.009$ & $33.745\pm0.010$ & $33.745\pm0.009$\\%
$P$ (d) & 69.728087 & 69.728107 & 69.728117 & \\
$\sigma_P$ (d) & 0.000014 & 0.000013 & 0.000014 & \\
$t_C - 2450000$ & 7180.14126 & 7180.14068 & 7180.14063 & \\
$\sigma_{t_C}$ & 0.00012 & 0.00012 & 0.00016 & \\
$i$ ($\degr$) & $89.565\pm0.019$ & $89.568\pm0.016$ & $89.637\pm0.013$ & \\
$q$ & $0.674\pm0.004$ & $0.674\pm0.004$ & $0.674\pm0.003$ & $0.674\pm0.004$\\
$e \cos \omega$ & $-0.036166\pm0.000009$ & $-0.036126\pm0.000010$ & $-0.036134\pm0.000005$ & \\
$e \sin \omega$ & $0.0377\pm0.0016$ & $0.0376\pm0.0016$ & $0.0361\pm0.0016$ & \\
$e$ & $0.0522\pm0.0012$ & $0.0522\pm0.0011$ & $0.0510\pm0.0011$ & \\
$\omega$ ($\degr$) & $133.7\pm1.1$ & $133.8\pm1.2$ & $135.1\pm1.3$ & \\
$K_1$ (km s$^{-1}$) & $25.64\pm0.04$ & $25.65\pm0.04$ & $25.63\pm0.04$ & $25.64\pm0.04$\\
$K_2$ (km s$^{-1}$) & $38.06\pm0.19$ & $38.06\pm0.20$ & $38.05\pm0.22$ & $38.06\pm0.20$\\
*$R_1/R_2$ & $1.447\pm0.042$ & $1.460\pm0.035$ & $1.537\pm0.020$ & \\
$R_1/a$ & $0.01221\pm0.00007$ & $0.01221\pm0.00008$ & $0.01225\pm0.000014$ & \\
$R_2/a$ & $0.00844\pm0.00025$ & $0.00836\pm0.00021$ & $0.00797\pm0.00016$ & \\
$T_2/T_1$ & $0.770\pm0.005$ & $0.772\pm0.006$ & $0.772\pm0.006$ & \\
K2 contam. & $0.031\pm0.004$ & $0.0027\pm0.0029$ & $0^{+0.0003}$ & \\
\hline
$L_2/L_1(B)$ & $0.0694\pm0.0049$ & $0.0686\pm0.0043$ & $0.0655\pm0.0028$ & $0.0690\pm0.0049\pm0.0004$(sys)\\
$L_2/L_1(V)$ & $0.1081\pm0.0067$ & $0.1076\pm0.0055$ & $0.0940\pm0.0025$ & $0.1079\pm0.0067\pm0.0003$(sys)\\
$L_2/L_1(R)$ & $0.1439\pm0.0090$ & $0.1420\pm0.0070$ & $0.1209\pm0.0033$ & $0.1429\pm0.0090\pm0.0010$(sys)\\
$L_2/L_1(I)$ & $0.1724\pm0.0099$ & $0.1734\pm0.0096$ & $0.1554\pm0.0052$ & $0.1729\pm0.0099\pm0.0005$(sys)\\
\hline
$M_1/\msun$ & $1.109\pm0.015$ & $1.112\pm0.013$ & $1.111\pm0.014$ & $1.111\pm0.015$\\
$M_2/\msun$ & $0.749\pm0.005$ & $0.749\pm0.005$ & $0.747\pm0.006$ & $0.748\pm0.005$\\
$R_1/\rsun$ & $1.070\pm0.008$ & $1.070\pm0.007$ & $1.073\pm0.006$ & $1.071\pm0.008\pm0.003$(sys) \\
$R_2/\rsun$ & $0.739\pm0.019$ & $0.732\pm0.016$ & $0.699\pm0.010$ & $0.713\pm0.019\pm0.026$(sys)\\
$\log g_1$ (cgs) & $4.425\pm0.006$ & $4.426\pm0.006$ & $4.422\pm0.005$ & $4.424\pm0.002$ \\
$\log g_2$ (cgs) & $4.575\pm0.021$ & $4.582\pm0.019$ & $4.624\pm0.011$ & $4.607\pm0.021\pm0.032$(sys)\\
\enddata
\end{deluxetable}

\end{document}